\documentclass{ismdproc}

\begin{document}
\title{PDFs for the LHC}

\author{{\slshape Amanda Cooper-Sarkar}  \\[1ex]
University of Oxford, Denys Wilkinson Bdg, Keble Rd, Oxford OX1 3RH, GB }


\maketitle

\begin{abstract}
  The PDF4LHC group has benchmarked six modern PDF sets and used them to make 
predictions for W,Z prdocution and Higgs production at the LHC. The reasons why 
predictions differ are examined and recent updates to the PDFs and their 
predictions are presented. 
\end{abstract}

\section{Introduction}

\begin{figure}[tbp]
\vspace{-1.0cm}
\begin{center}
\begin{tabular}{ll}
\psfig{figure=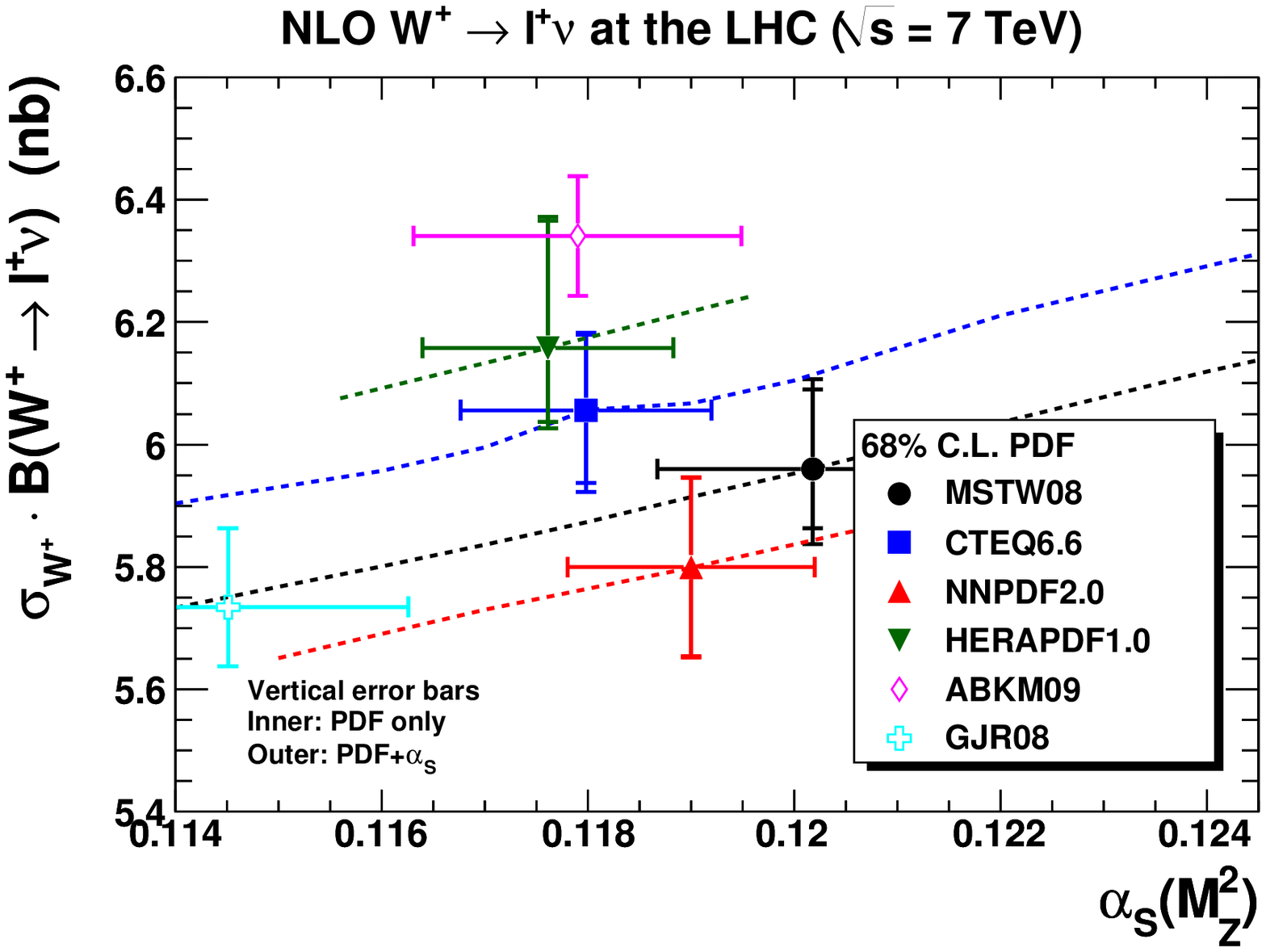,width=0.40\textwidth}~~
\psfig{figure=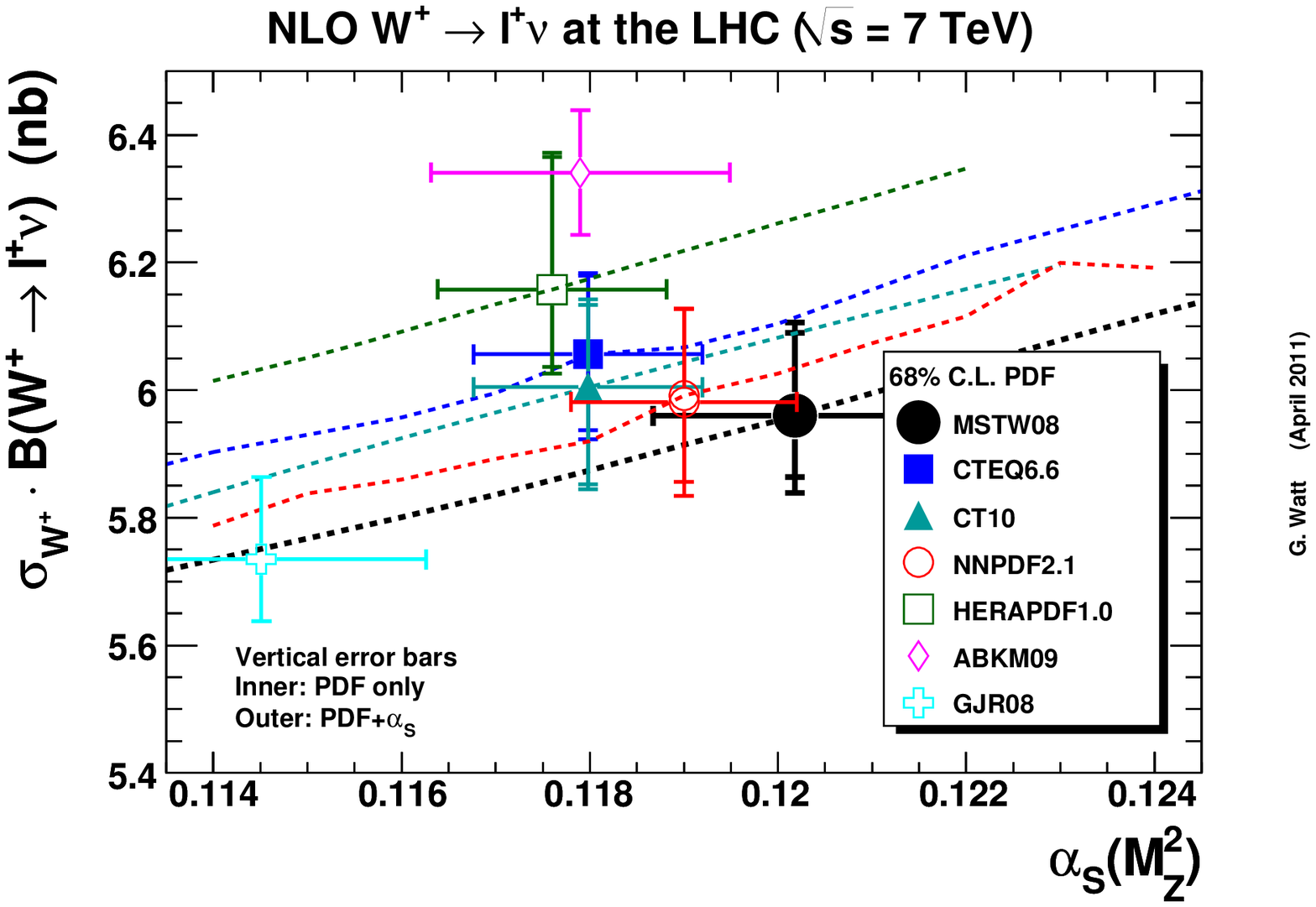,width=0.40\textwidth} \\
\end{tabular} 
\caption{Predictions for the $W+$ cross section at the LHC at 7 TeV from 
various modern PDF sets as a function of $\alpha_s(M_Z)$. The cross section 
for each PDF set is shown at the value of $\alpha_s(M_Z)$ used by that group.
The vertical error bars represent the $68\%$ uncertainty on the predictions, 
the 
horizontal error bars represent the $68\%$ uncertainty on $\alpha_s(M_Z)$ 
considered by each group. 
The right hand plot shows the predictions for PDF sets available in April 2010,
the left hand side shows the updates from CT10 and NNPDF2.1 
available for April 2011. Plots from G.Watt http://projects.hepforge.org/mstwpdf/pdf4lhc/2010/
}
\label{fig:W+}
\end{center}
\end{figure}

The PDF4LHC group have benchmarked modern PDF sets from the groups; ABKM, CTEQ,
GJR, HERAPDF, NNPDF, MSTW in terms of the predictions for basic LHC 
cross-sections~\cite{Alekhin:2011sk}. Fig~\ref{fig:W+} shows the comparison of 
predictions for the $W^+$ cross section at 7 TeV. The PDF4LHC 
recommendation~\cite{Botje:2011sn}
 is to use the envelope of the MSTW2008, 
CTEQ6.6 and NNPDF2.0 predictions and add the uncertainty due to 
$\alpha_s(M_Z)$ in quadrature. However, as can be seen from the figure, 
this may not be sufficient to cover PDF uncertainties.

There are several reasons why the PDF predictions differ: firstly they are 
based on different data sets. For example, all the 2010 PDF analyses 
illustrated, bar the HERAPDF1.0 and the NNPDF2.0, do 
not use the recently combined inclusive cross section data from 
HERA-I~\cite{h1zeus:2009wt} which are up to three times more accurate than the 
separate H1 and ZEUS data sets used by previous PDF analsyses. 
These combined HERA data are also shifted in normalisation by $\sim 3\%$ with 
respect to the previous HERA data, and this partly explains the higher cross 
section of the HERAPDF wrt CTEQ and MSTW. Secondly, PDFs use different central 
values of $\alpha_s(M_Z)$, the effect of this is illusrated on the figure.
Some groups (HERAPDF, CTEQ, NNPDF) adopt a central value of $\alpha_S(M_Z)$ 
inspired by the PDG value and others (ABKM, GJR, MSTW) fit $\alpha_s(M_Z)$ 
simultaneously with 
the PDF parameters and use their best fit value. Thirdly, the PDF analyses 
differ in the schemes 
used to account for heavy quark production. For 
example, the NNPDF2.0 used a zero-mass variable-flavour number scheme (ZMVFN) 
and this explains why the NNPDF2.0 predictions lie lower than CTEQ, MSTW, 
HERAPDF all of which use general mass variable flavour number schemes (GMVFNs).
The value of the charm and beauty masses also differ between the PDF analyses.
Lastly, PDFs differ regarding choices of  PDF parametrisation and 
theoretical/model prejudices which are imposed. Whereas this latter source of 
difference 
represents legitimate, irreducible differences in approach it may be possible 
to achieve greater concordance on the first three reasons for PDF differences.

\section{Heavy Quark Schemes}
\begin{figure}[tbp]
\vspace{-3.0cm} 
\centerline{\psfig{figure=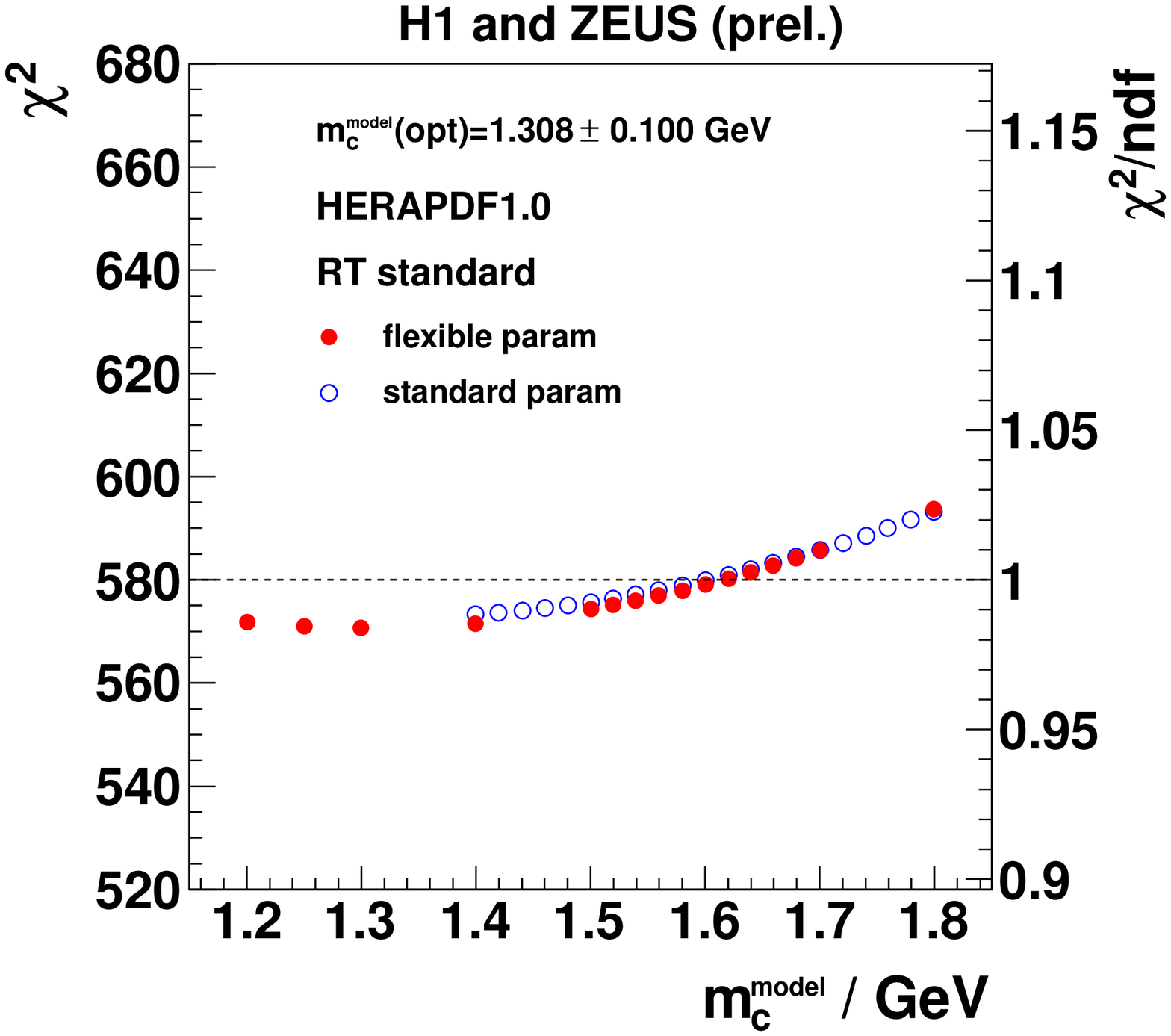,width=0.30\textwidth}~~ 
\psfig{figure=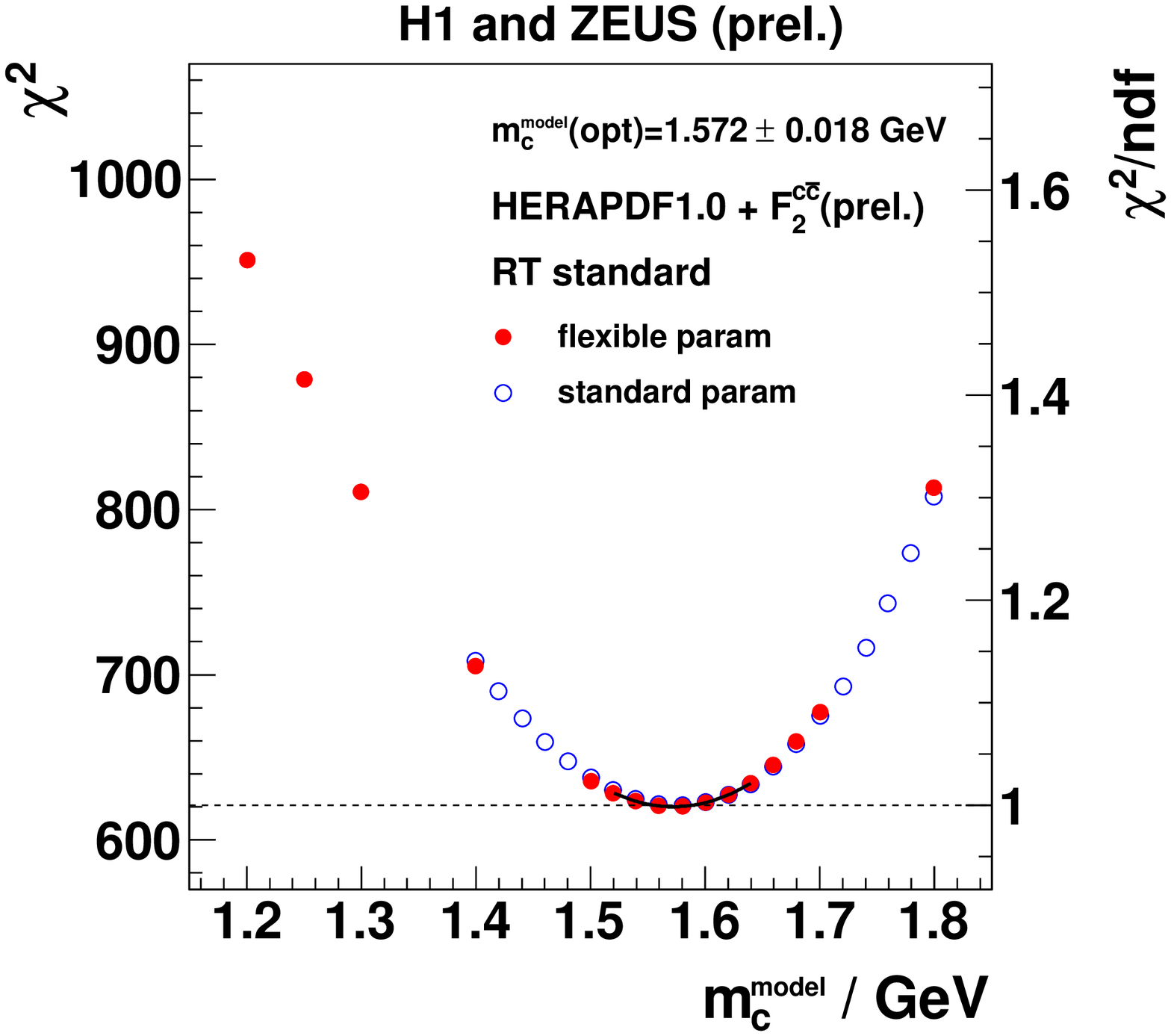,width=0.30\textwidth}}
\centerline{\psfig{figure=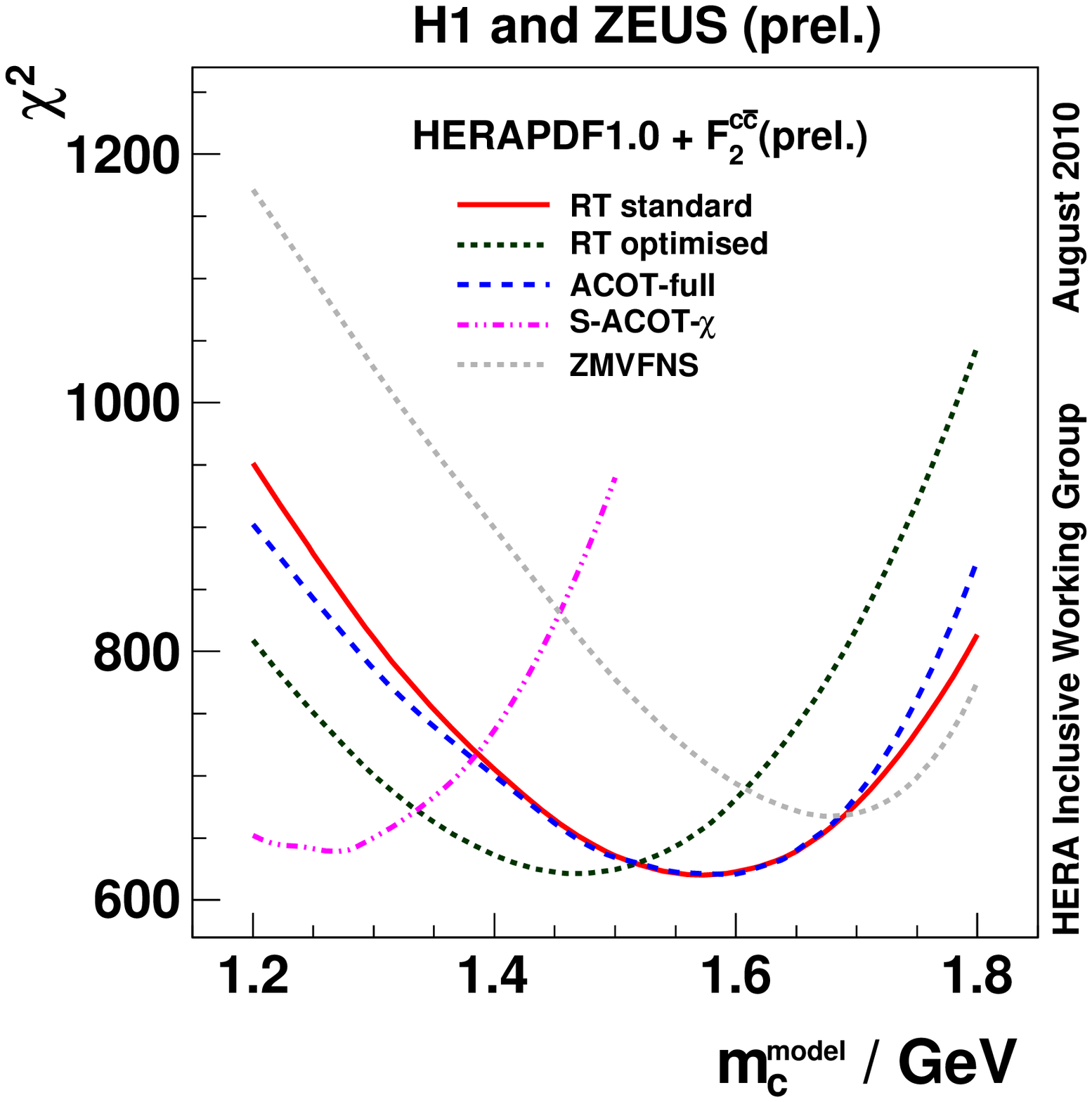,width=0.30\textwidth}~~ 
\psfig{figure=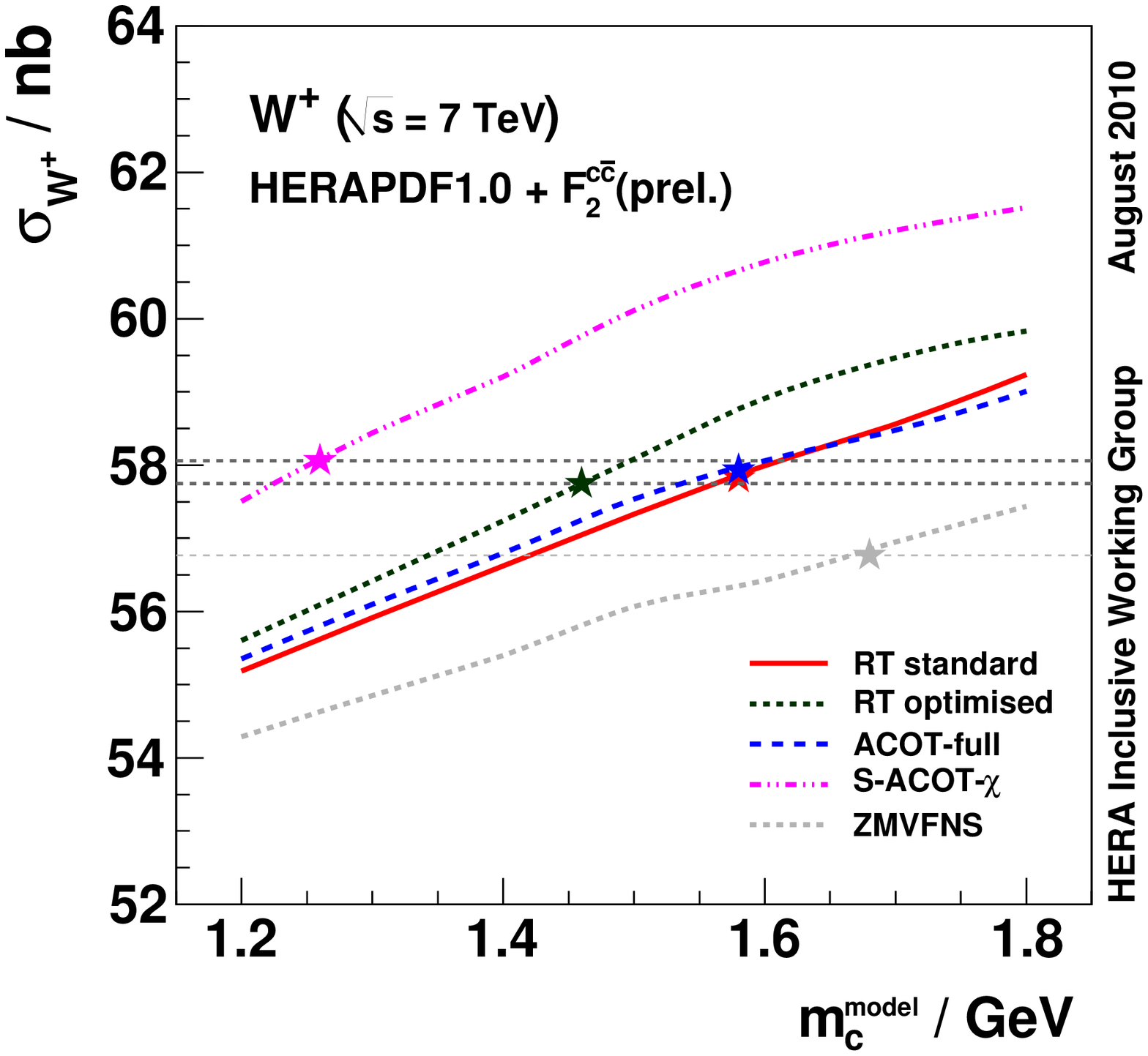,width=0.30\textwidth}}
\caption {The $\chi^2$ of the HERAPDF fit as a 
function of the charm mass parameter $m_c^{model}$. Top left; using the 
RT-standard heavy-quark-mass scheme, when only inclusive DIS data are included 
in the fit. Top right; 
using the RT-standard heavy-quark-mass scheme, 
when the data for $F_2^{c\bar{c}}$ are also included in the fit. Bottom left;
 using various heavy-quark-mass schemes, when the data for $F_2^{c\bar{c}}$
 are also included in the fit. Bottom right: predictions for the $W^+$ cross-sections at the LHC, as a 
function of the charm mass parameter $m_c^{model}$, for various heavy-quark-mass schemes.  
}
\label{fig:charmpred}
\end{figure}

Let us first consider heavy quark production. 
The ABKM and GJR groups use 
Fixed Flavour Number (FFN) treatments, HERAPDF, CTEQ and MSTW use GMVFN and 
NNPDF2.0 used ZMVFN. However even within GMVFN schemes there is not complete 
agreement. Predictions for $F_2^c$ differ 
between schemes~\cite{Rojo:2010gv} and the choice of scale within a scheme affects 
predictions. The value of the charm mass can also affect 
predictions, HERAPDF, NNPDF and MSTW now provide PDFs at different charm mass 
values so that the effect of this can be evaluated.
H1 and ZEUS have recently combined their data on 
$F_2^{c\bar{c}}$~\cite{charmcomb}, and these data can help to reduce the 
uncertainty on PDFs coming from the choice of scheme and the value of the 
charm mass.
These data have been input to the HERAPDF fit together with the inclusive data 
which were used for HERAPDF1.0. The $\chi^2$ of this fit is sensitive to the 
value of the charm quark mass. Fig.~\ref{fig:charmpred} compares the $\chi^2$, 
as a function of this mass, for a fit which includes these data (top right)
 to that for the HERAPDF1.0 fit (top left). 
However, it would be premature to conclude that the data 
can be used to determine the charm pole-mass. The HERAPDF formalism uses 
the Thorne-Roberts (RT) variable-flavour-number (VFN) scheme for heavy 
quarks. This scheme is not unique,
specific choices are made for threshold behaviour. In Fig.~\ref{fig:charmpred} 
(bottom left) the $\chi^2$ profiles for the standard and the 
optimized versions (optimized for smooth 
threshold behaviour~\cite{Thorne:2010pa}) of this scheme are compared. The same figure also
 compares the alternative ACOT VFN 
schemes and the Zero-Mass VFN scheme. Each of theese schemes 
favours a different value for the charm quark mass, and the fit to the data 
is equally good for all the heavy-quark-mass schemes excepting 
the zero-mass scheme. 
Each of these schemes can also be used to predict $W$ 
and $Z$ production for the LHC and their predictions for $W^+$ are shown in 
Fig.~\ref{fig:charmpred} as a function of the charm quark mass (bottom right). 
If a particular
value of the charm mass is chosen then the spread of predictions is as large as
$\sim7\%$.
However this spread is considerably reduced $\sim 1\%$ if each heavy quark 
scheme is used at its own favoured
value of the charm quark-mass.  Furher details of this study are 
given in ref.~\cite{charmstudy}.

A further relevant question is what is the charm mass which is being used in 
the calculations? Is it the pole mass or the running mass? The running mass is 
independently measured and the ABM group have considered using this mass in 
their FFN calculations~\cite{Alekhin:2010sv}. They obtain a value of $m_c(m_c) = 1.18\pm 0.06$ GeV 
in agreement with the PDG value. 
This should be improved after the input of the combined HERA charm data.

\section{2011 updates}
\begin{figure}[tbp]
\vspace{-2.0cm} 
\begin{center}
\begin{tabular}{ll}
\psfig{figure=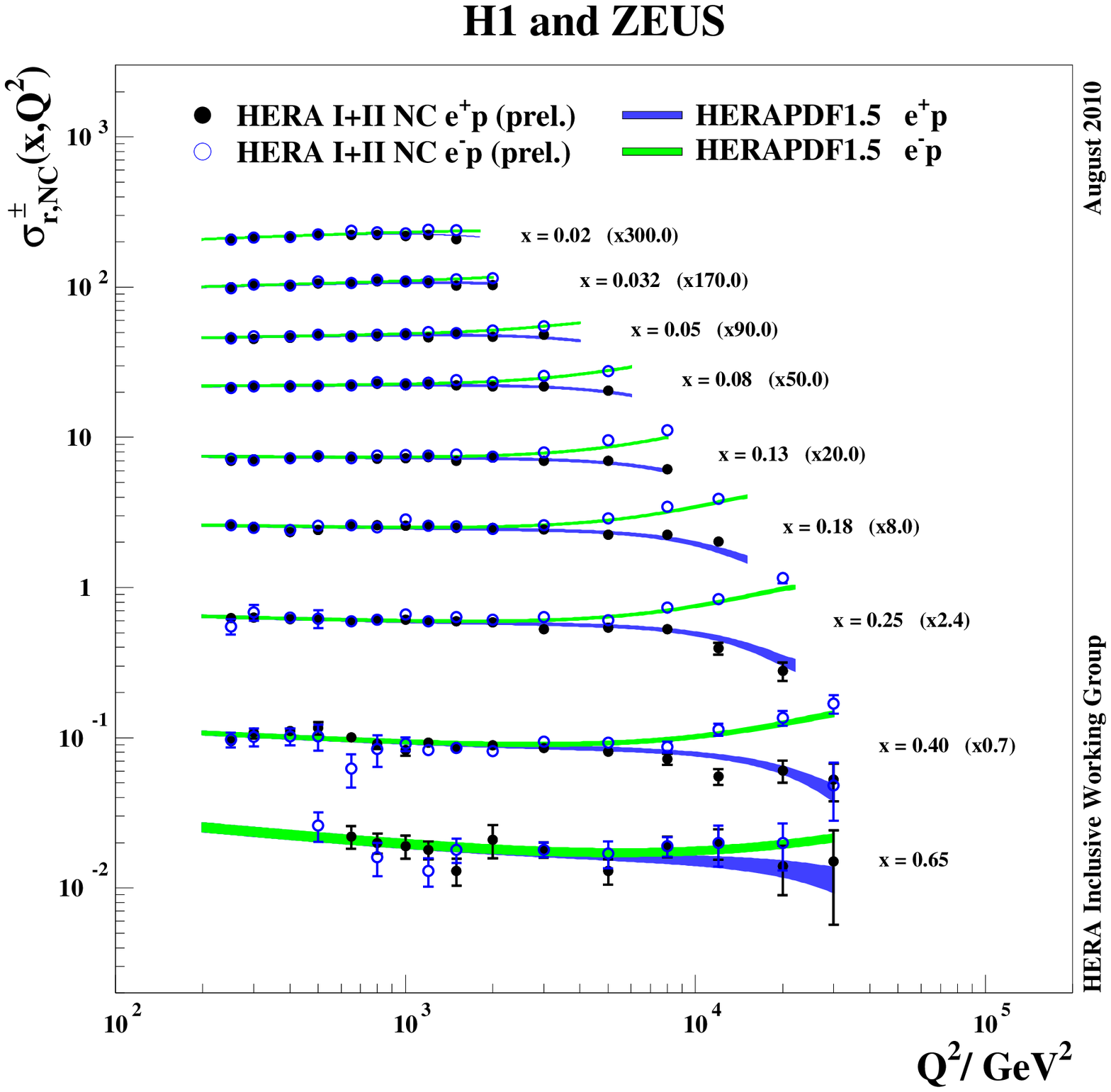,width=0.45\textwidth}~~ &
\psfig{figure=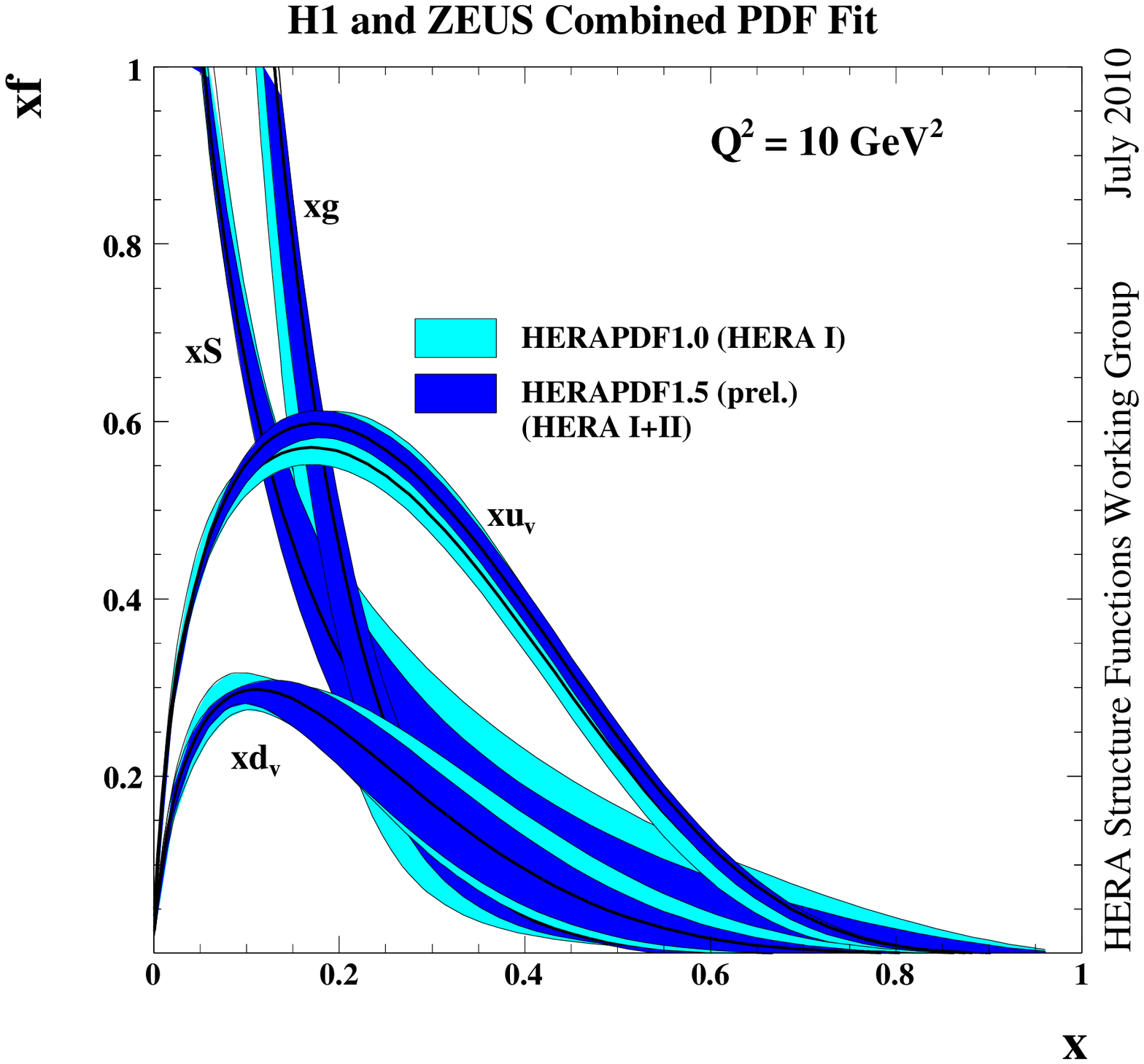,width=0.45\textwidth} \\
\end{tabular} 
\caption{Left: HERA combined data points for the NC $e^{\pm}p$ cross-sections 
as a function of $Q^2$ in bins of $x$, for data from the HERA-I and II run 
periods. The HERAPDF1.5 fit to these data is also shown on the plot.
Right: Parton distribution functions from HERAPDF1.0 and HERAPDF1.5; $xu_v$, 
$xd_v$,$xS=2x(\bar{U}+\bar{D})$ and $xg$ at $Q^2=10~$GeV$^2$.}
\label{fig:herapdf15}
\end{center}
\end{figure}

There have been recent updates to the PDFs which address some of the issues we have raised.
NNPDF2.1~\cite{Rojo:2010gv}  updates NNPDF2.0 to use a GMVFN. CT10~\cite{Lai:2010vv} is an update of CTEQ6.6 which 
includes the combined HERA data in addition to giving special consideration to the 
Tevatron W asymmetry data (CT10W). ABM11~\cite{Alekhin:2011cf} is an update of ABKM which includes 
the combined HERA data. HERAPDF1.0 has been updated to HERAPDF1.5~\cite{hiq2study} by including 
further preliminary combined data from HERA-II running.
Preliminary H1 data on $NC$ and $CC$ $e^+p$ and $e^-p$ inclusive cross-sections
and published ZEUS data on $NC$ and $CC$ $e^-p$ and $CC$ $e^+p$ data, from 
HERA-II running, have been combined with the HERA-I data to yield an inclusive 
data set wih improved accuracy at high $Q^2$ and high $x$~\cite{highq2}. 
 This new data set is used as the 
sole input to  a PDF fit called HERAPDF1.5 
which uses the same formalism 
and assumptions as the HERAPDF1.0 fit. 
Fig.~\ref{fig:herapdf15} (left) shows the combined data for $NC$ $e^{\pm}p$ 
cross-sections with the HERAPDF1.5 fit superimposed. The parton distribution
functions from HERAPDF1.0 and HERAPDF1.5 are compared in 
Fig.~\ref{fig:herapdf15} (right). The improvement in precision at high $x$ 
is clearly visible.

Fig.~\ref{fig:W+}, right hand side, shows the predictions fot the W+ cross 
section at the 
LHC at 7 TeV updated to show the predictions of the new PDFs from CT10 
and NNPDF2.1. 
The CT10 and CTEQ6.6 predictions 
are very similar and the HERAPDF1.5 (not shown) and HERAPDF1.0 predictions are very 
similar. The NNPDF2.1 prediction has moved up significantly beacuse of the use
 of the GMVFN scheme. The early LHC data agree well with all the predictions. 

\begin{figure}[tbp]
\vspace{-1.0cm} 
\centerline{\psfig{figure=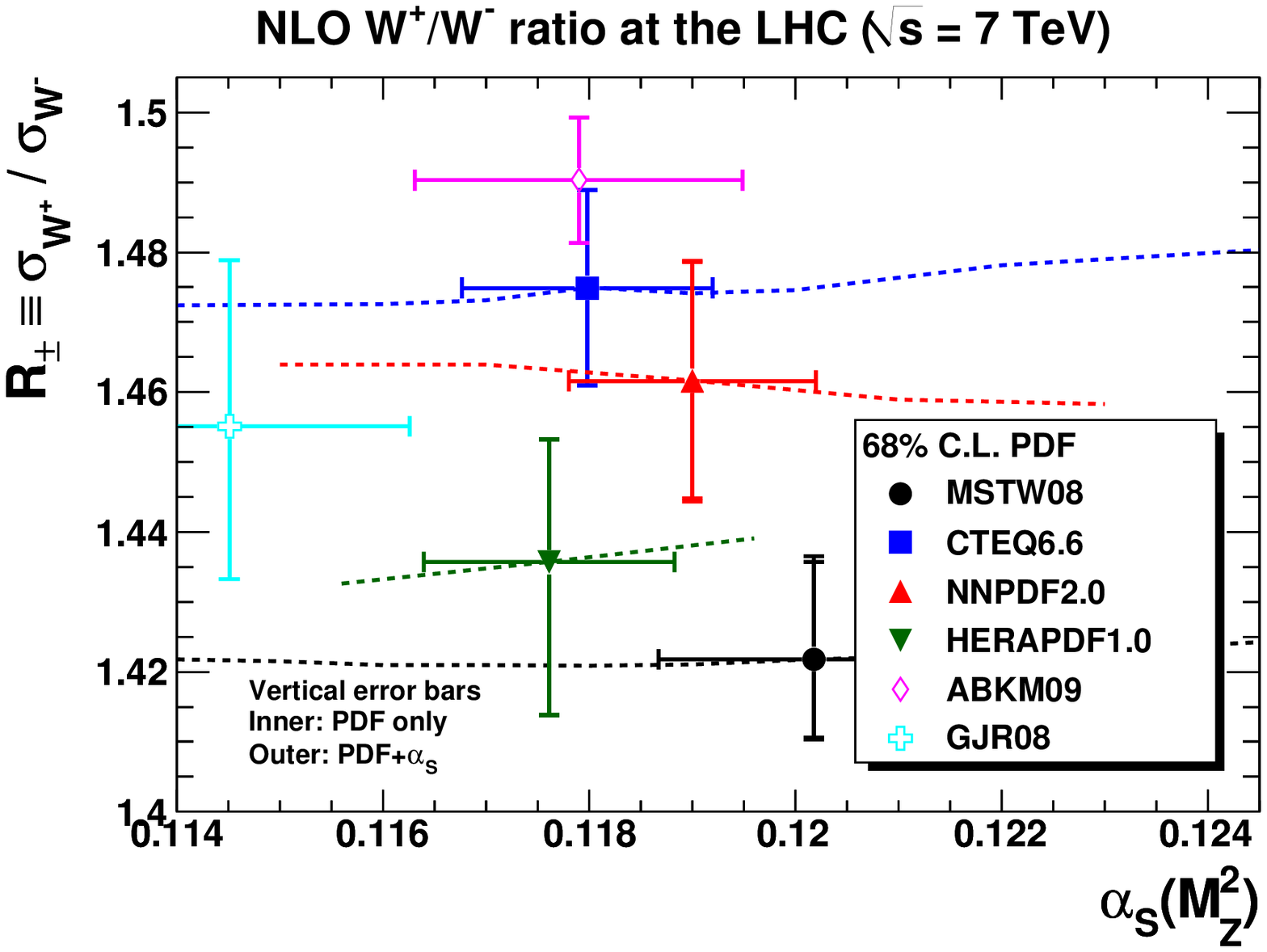,width=0.30\textwidth}~~ 
\psfig{figure=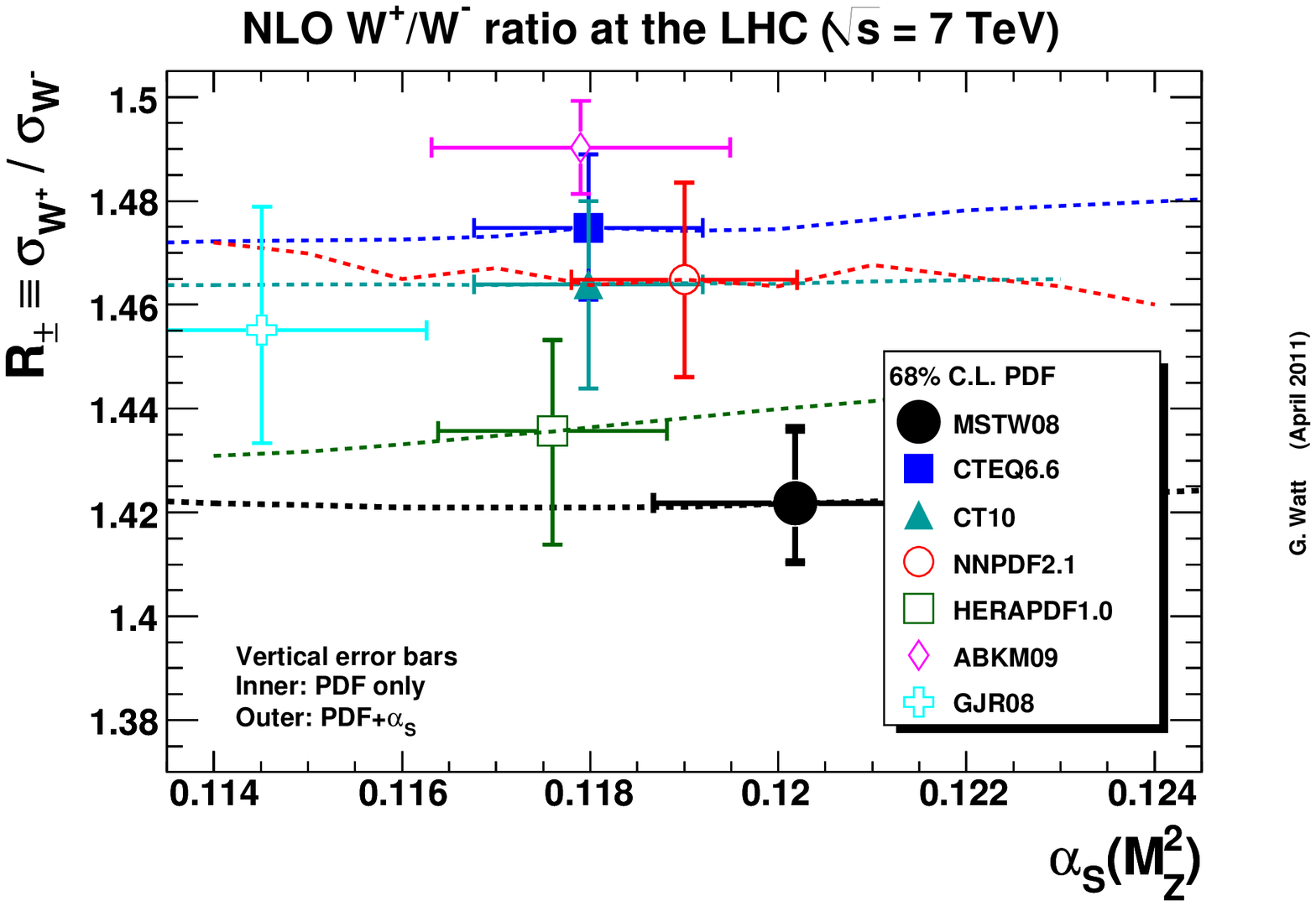,width=0.30\textwidth}}
\centerline{\psfig{figure=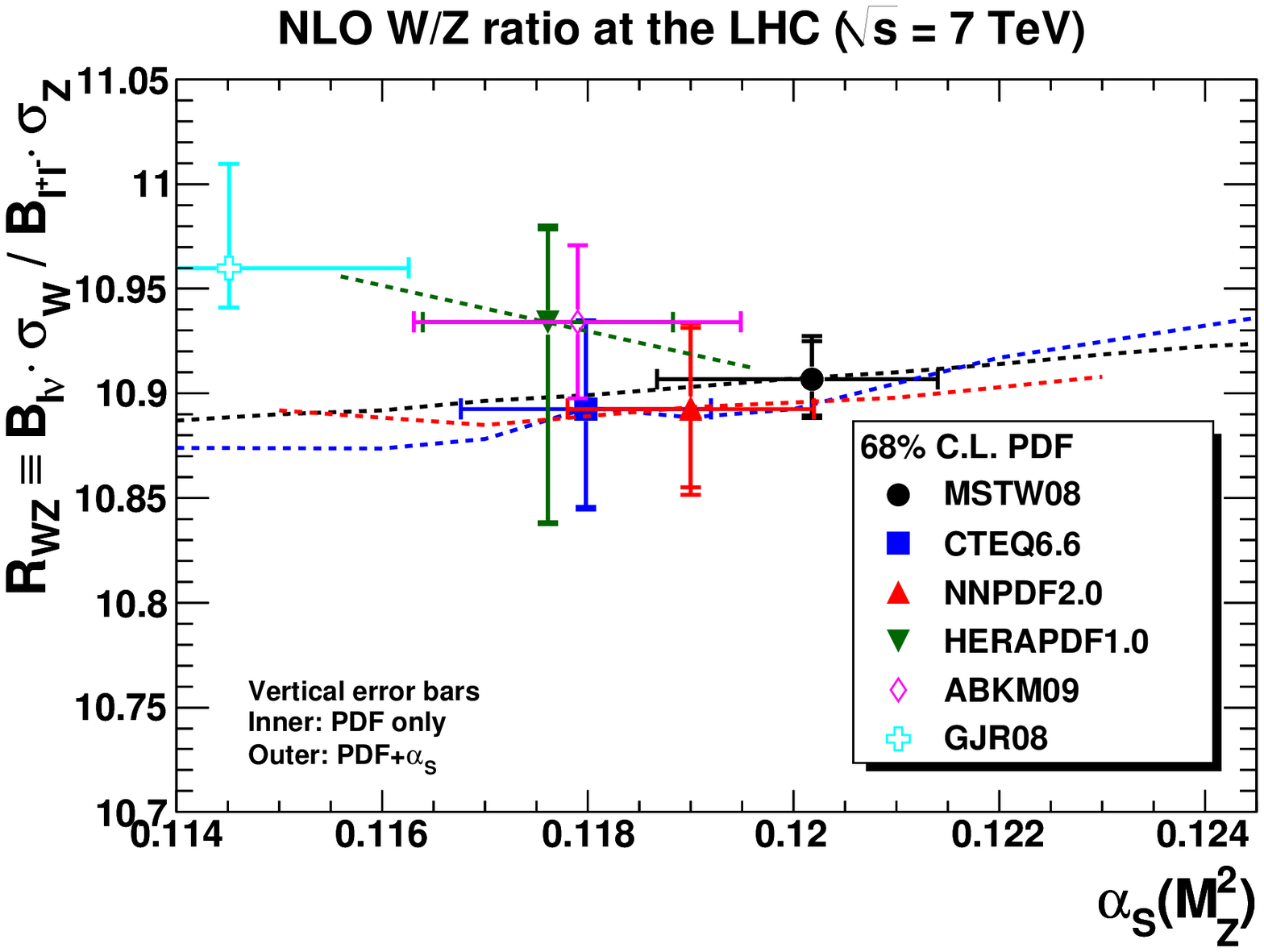,width=0.30\textwidth}~~ 
\psfig{figure=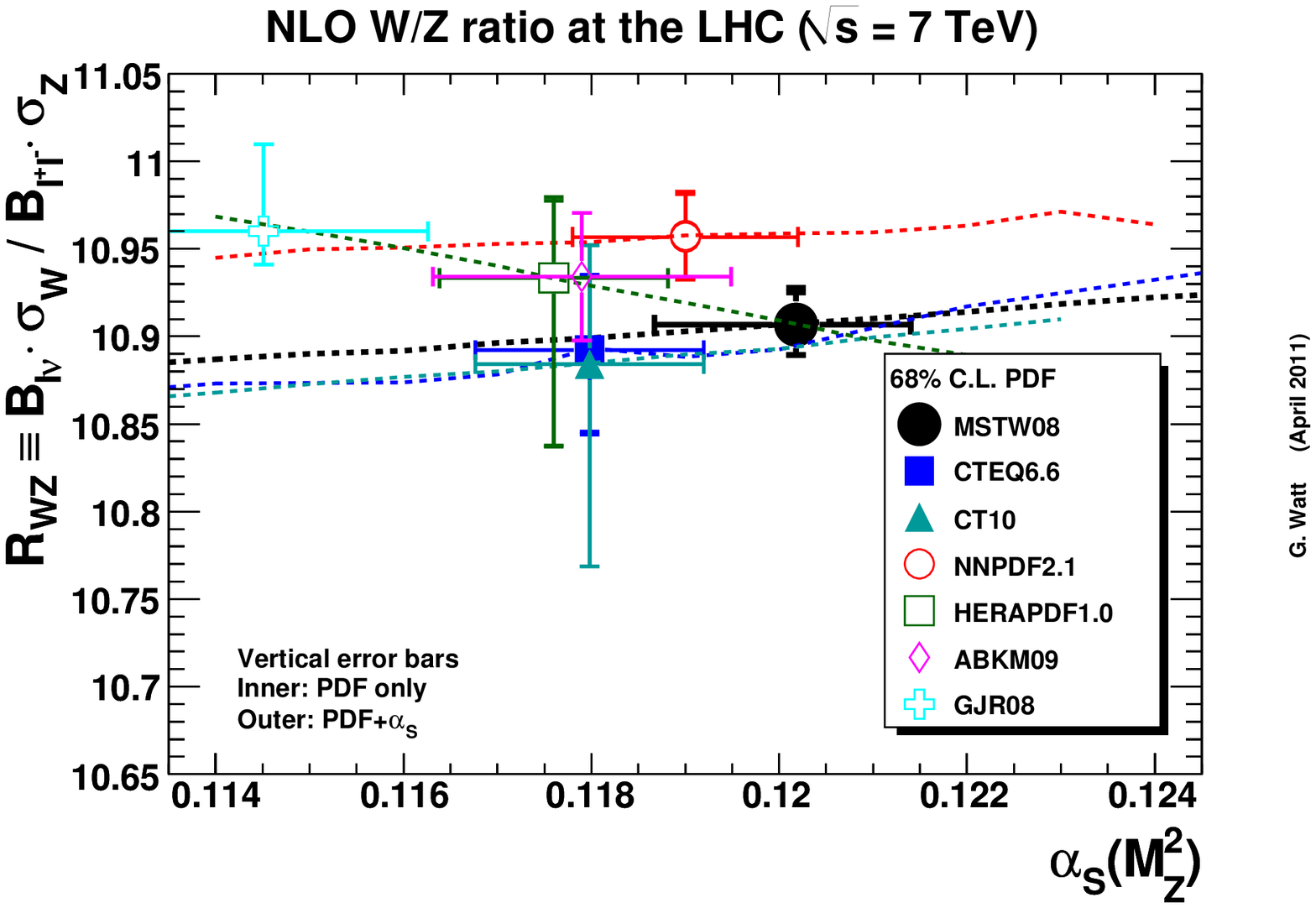,width=0.30\textwidth}}
\caption {Upper plots: the ratio of $W^+$ and $W^-$ cross sections at the LHC 
for the PDFs considered by the PDF4LHC group. 
Lower plots: the ratio of $Z$ to $W^+ + W^-$ 
cross sections at the LHC 
for the PDFs considered by the PDF4LHC group. Left-hand side April 2010, 
right-hand side April 2011. Plots from G.Watt http://projects.hepforge.org/mstwpdf/pdf4lhc/2010/
}
\label{fig:ratios}
\end{figure}

It is also interesting to look at the predictions for the ratios of cross 
sections $Z/(W^+ + W^-)$ and $W^-/W^+$. Fig.~\ref{fig:ratios} shows 2010 and 
updated 2011 predictions. The $Z/W$ ratio is very consistently predicted because
flavour dependence almost cancels in the ratio, but the $W^+/W^-$ ratio 
predictions are quite discrepant - this measures a difference in the $u$ and 
$d$ valence distributions in a previously unmeasured region of $x$. 
The early LHC data are not yet discriminating.

\begin{figure}[tbp]
\vspace{-1.0cm} 
\centerline{\psfig{figure=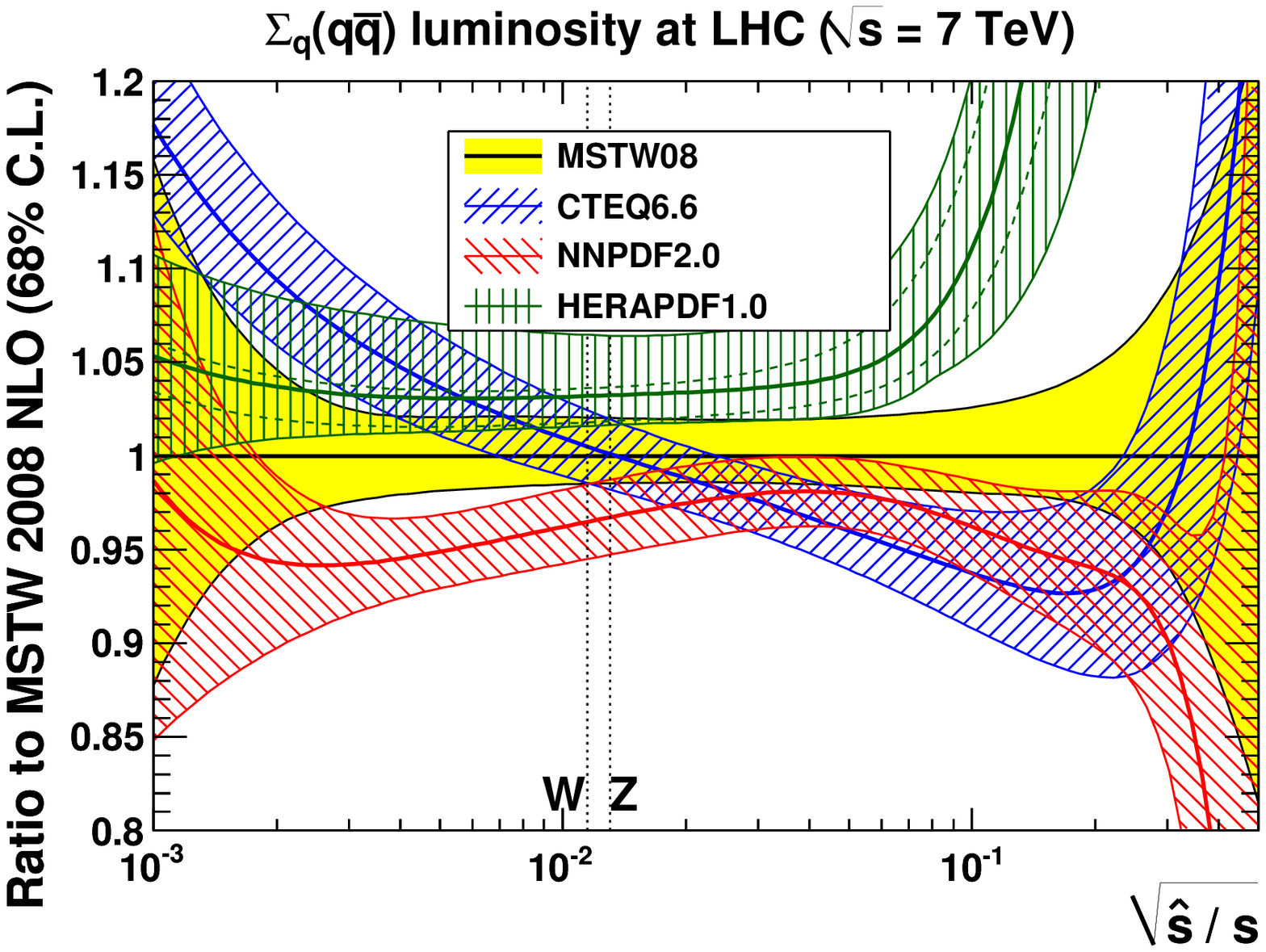,width=0.30\textwidth}~~ 
\psfig{figure=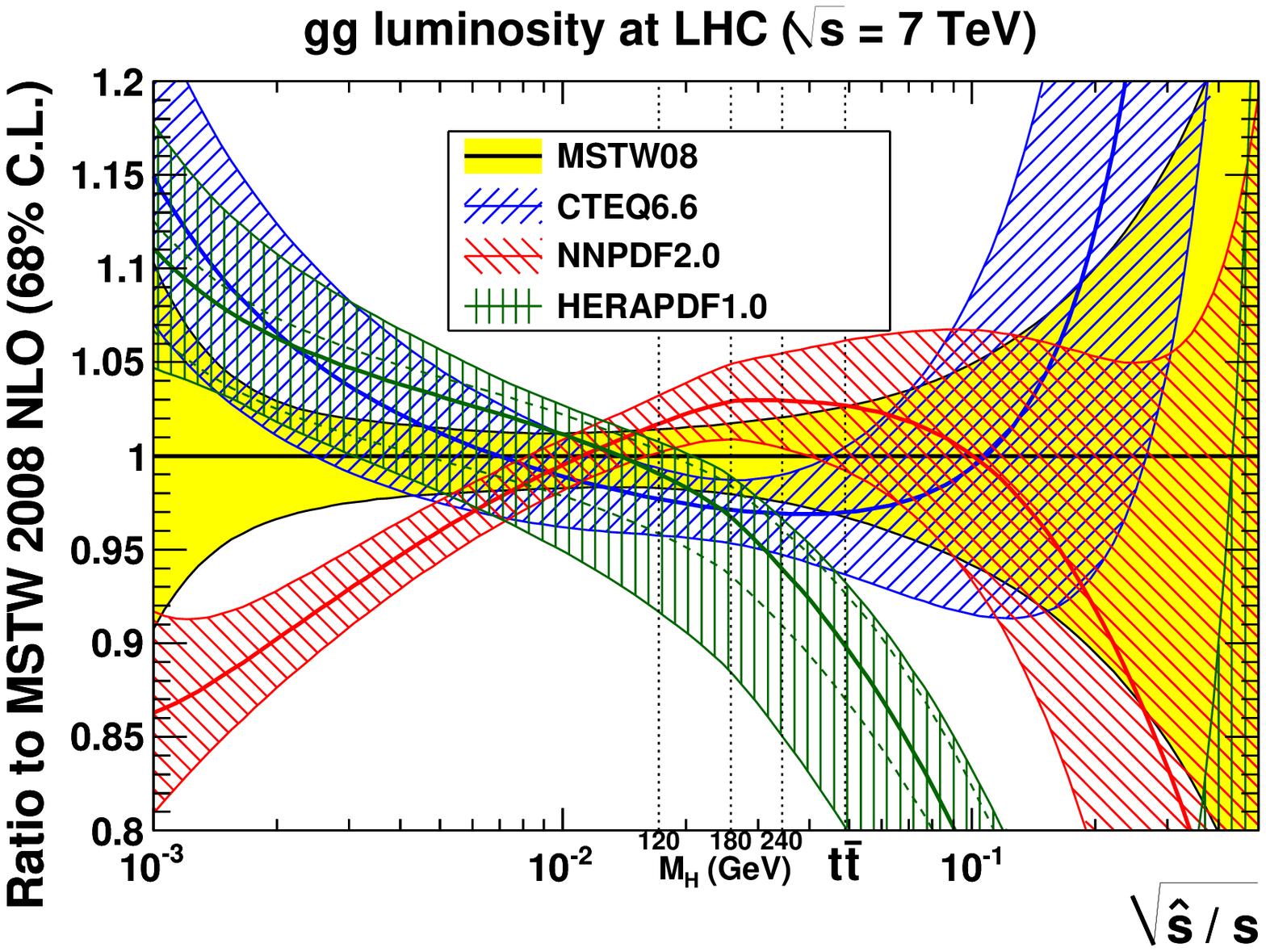,width=0.30\textwidth}}
\centerline{\psfig{figure=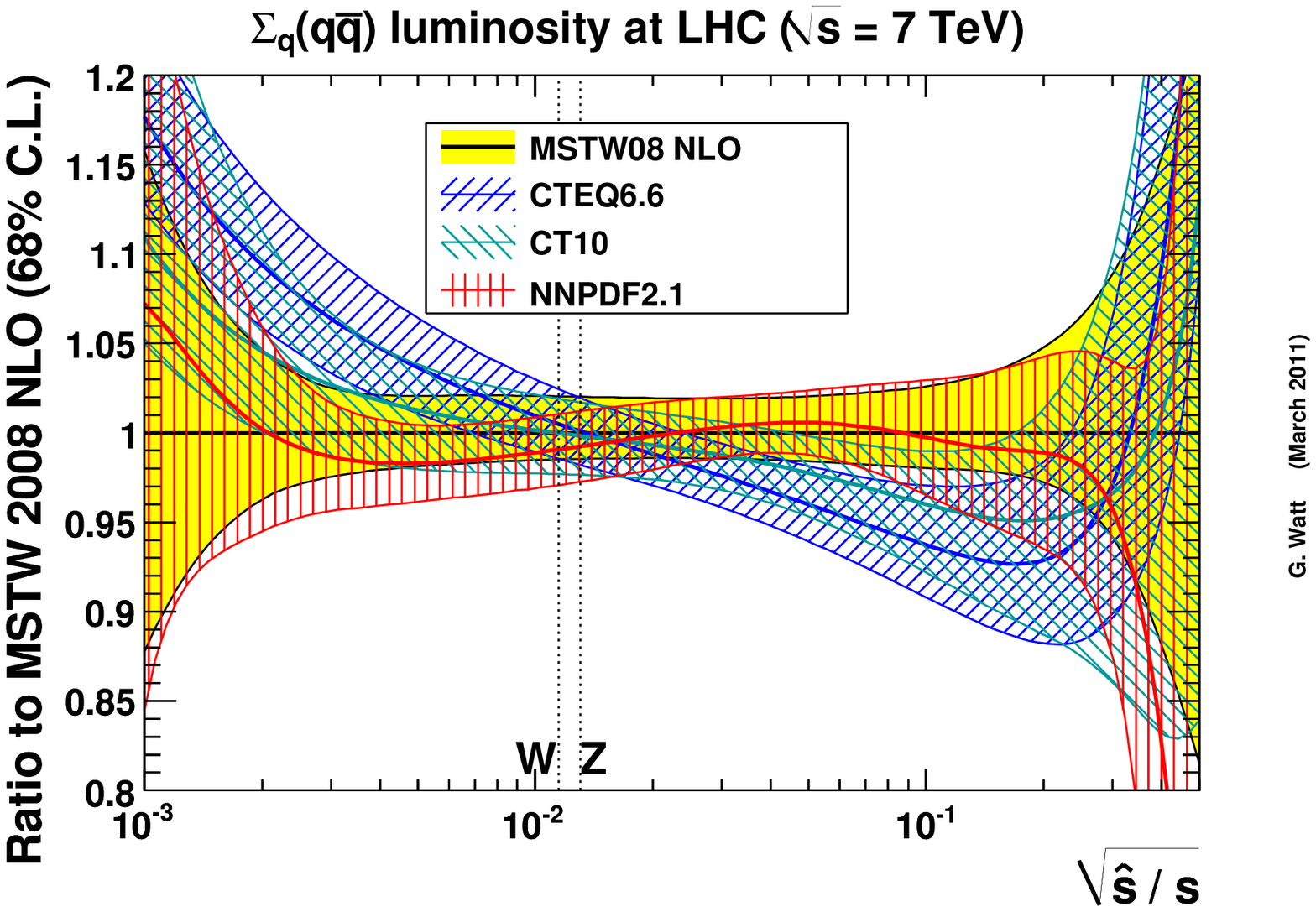,width=0.30\textwidth}~~ 
\psfig{figure=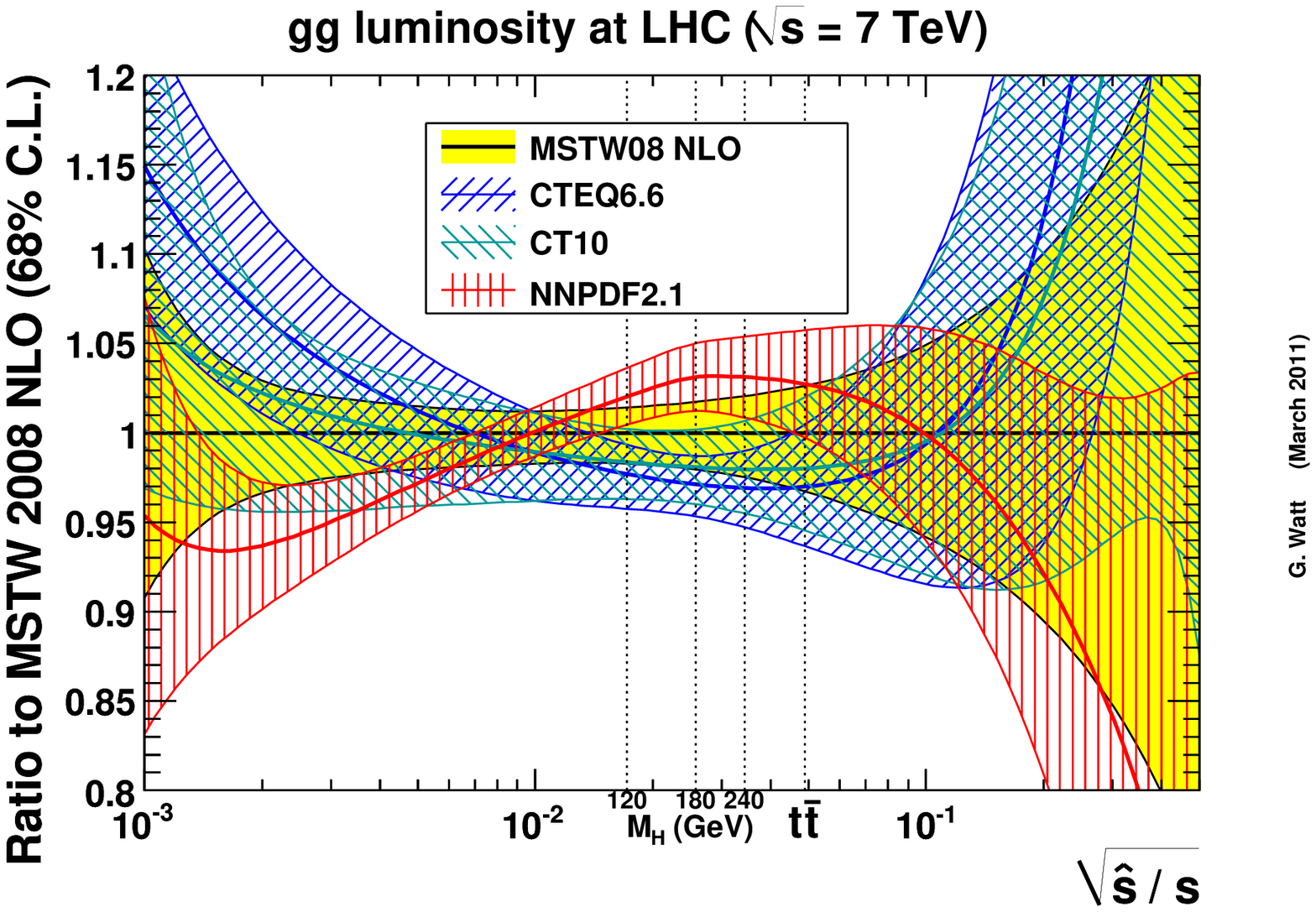,width=0.30\textwidth}}
\caption {Left hand side: the $q-\bar{q}$ luminosity in ratio to that of MSTW2008 for
 various PDFs. Right hand side: the 
same for the $g-g$ luminosities.
Upper row, PDFs 2010: bottom row, updates for CT10 and NNPDF2.1 2011.
Plots from G.Watt http://projects.hepforge.org/mstwpdf/pdf4lhc/2010/}
\label{fig:lumi}
\end{figure}

Another way to compare PDF predictions is  to look at parton-parton 
luminosities. Fig.~\ref{fig:lumi} shows the 2010 and 2011 $q-\bar{q}$ and $g-g$
luminosity plots. The 2011 comparisons are is considerably better agreement.

\section{The value of $\alpha_s(M_Z)$}

\begin{figure}[tbp]
\vspace{-2.0cm} 
\centerline{\psfig{figure=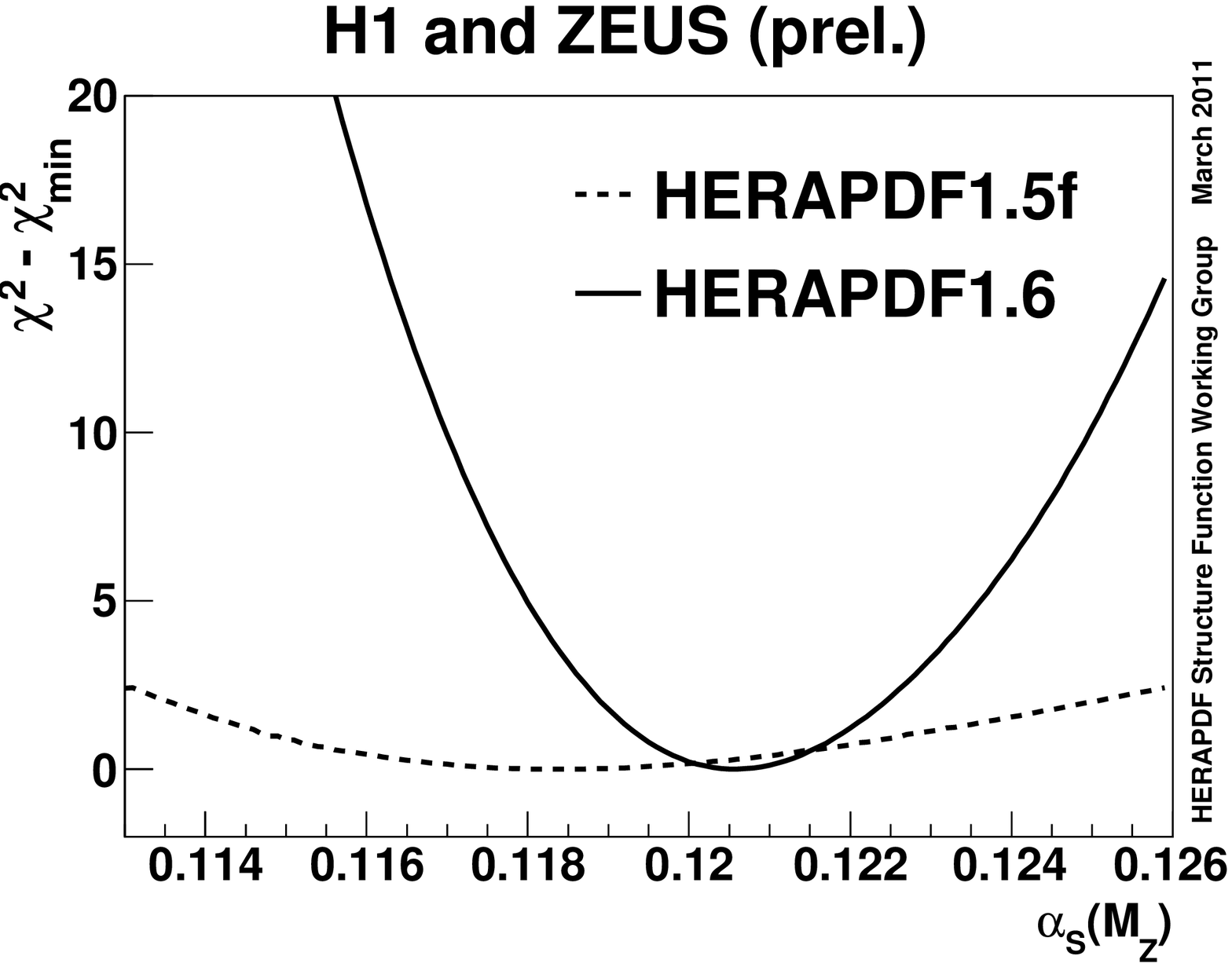,width=0.30\textwidth}~~ 
\psfig{figure=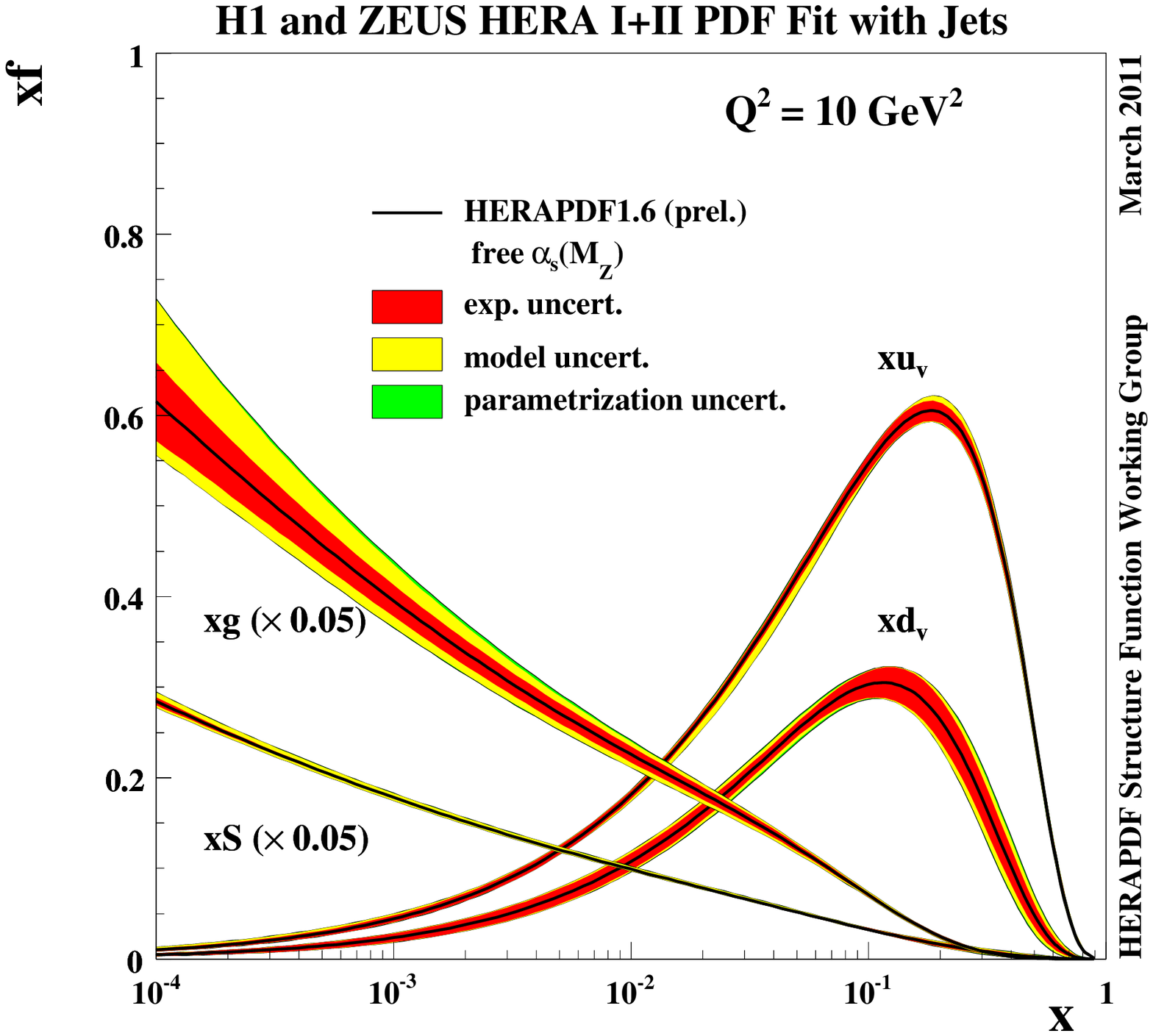,width=0.30\textwidth}}
\centerline{\psfig{figure=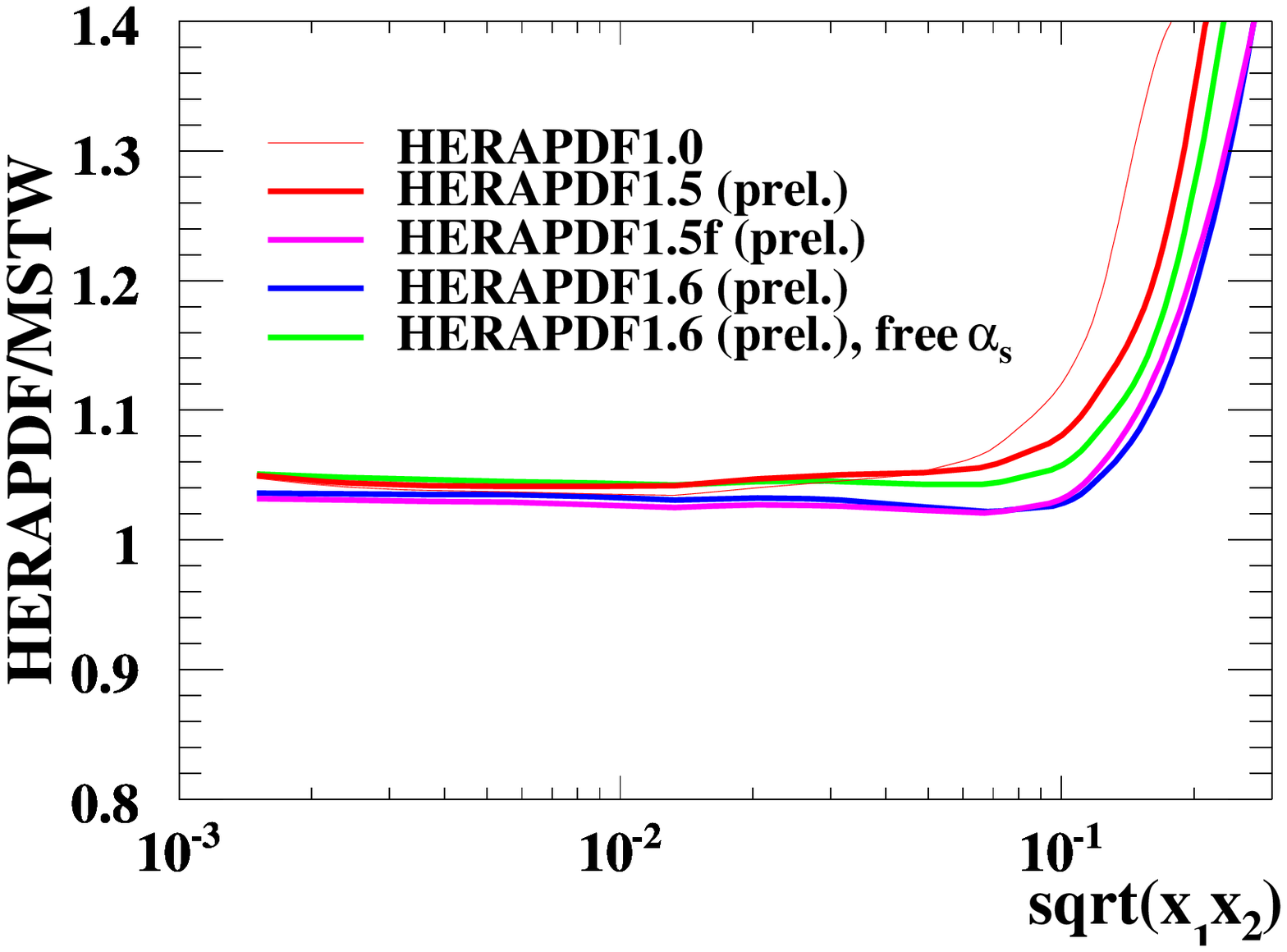,width=0.30\textwidth}~~ 
\psfig{figure=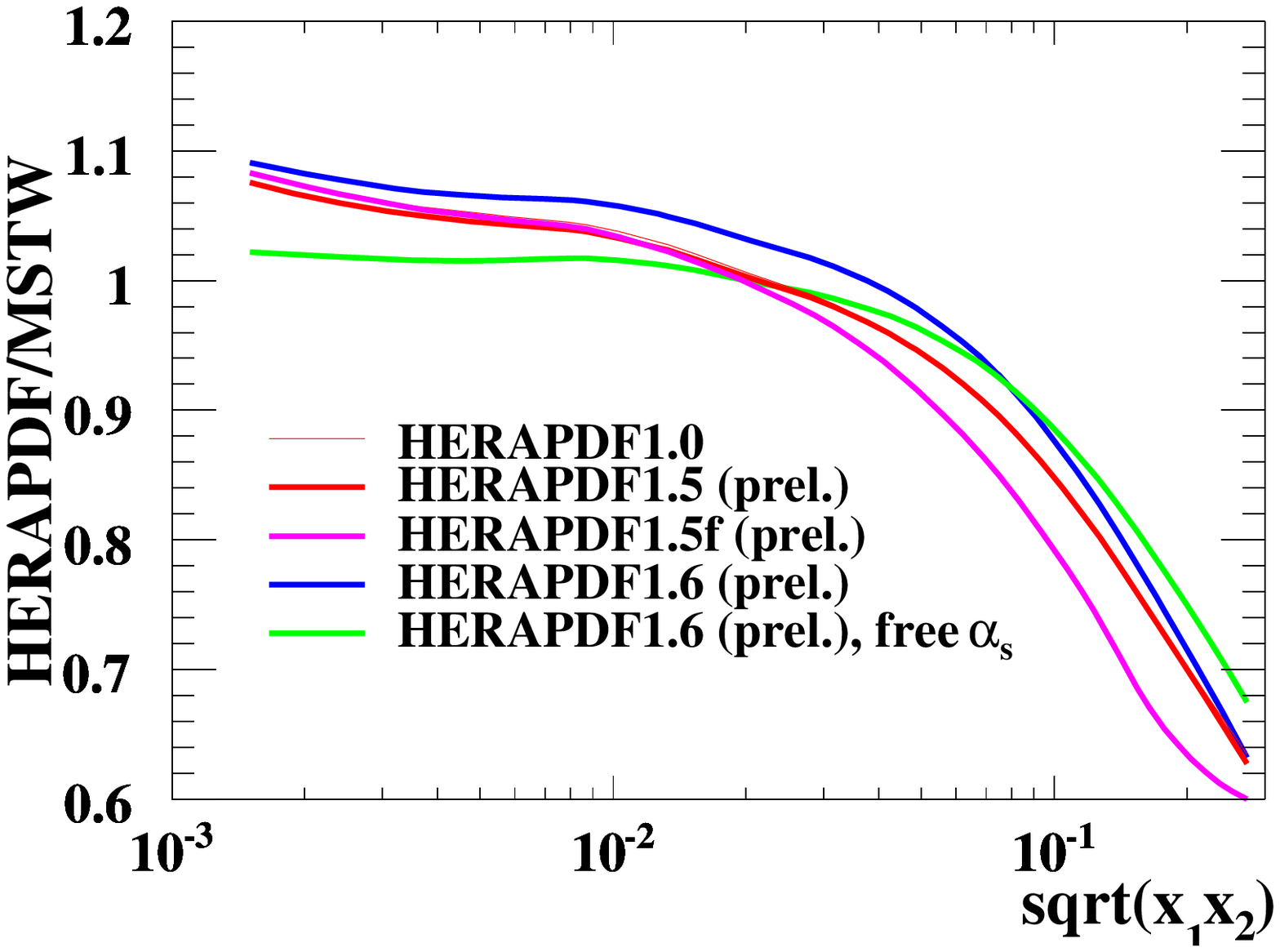,width=0.30\textwidth}}
\caption {Top left: $\chi^2$ scan vs $\alpha_s(M_Z)$ for the HERAPDF1.5f fit 
without jet data and for the HERAPDF1.6 fit which uses HERA jet data. 
Top right the PDFs resulting from the HERAPDF1.6 fit with 
free $\alpha_s(M_Z)$. Bottom row the $q-\bar{q}$ (left) and $g-g$ (right) luminosities in ratio to MSTW2008 for the 2011 HERAPDF updates and for HERAPDF1.0.
}
\label{fig:chi2prof}
\end{figure}

All groups bar GJR use values 
$\sim 0.118-0.120$ at NLO but there is a definite low($0.113$)- high($0.117$) 
split at NNLO. MSTW obtain the highest value at both NLO and NNLO and this 
has been 
atributed to the use of Tevatron jet data in their fits. However, ABM have 
tried inputting these data and find this has only a small effect on their 
$\alpha_s(M_Z)$ extraction~\cite{Alekhin:2011cf}. There is also a 'folk-lore' that DIS data prefer 
lower values of $\alpha_s$ however both MSTW~\cite{Martin:2009bu} and NNPDF~\cite{Lionetti:2011pw} have performed DIS only 
fits in which they find that only the BCDMS data prefer low $\alpha_s$ values. 
The HERA data actually prefer quite high values. This year we have heard from 
HERA themselves~\cite{herapdf16}. HERA have input jet data from H1 and ZEUS 
in addition to the HERA inclusive DIS data to obtain the HERAPDF1.6 fit. 
This fit also 
extends the HERAPDF1.5 parametrisation to use 14 free parameters 
(this fit is called HERAPDF1.5f)- a term is 
added such that the gluon may become negative at low $x,Q^2$ if required 
(though it does not do so in the kinematuc region where data is fitted) and 
the low-$x$ valence shape of the d-quark is freed from that of the u-quark. 
When $\alpha_S$ 
is freed a value of $\alpha_s(M_Z)=0.1202\pm 0.0019$ is 
obtained (where the error excludes the scale 
errors). This fit also has a harder high-x gluon density 
(and corresponding lower low-x gluon density) than the HERAPDF1.5 fit, 
which brings the gluon-gluon 
luminosity plot for HERAPDF1.6 into closer agreement with MSTW2008. 
Fig.~\ref{fig:chi2prof} 
shows the $\chi^2$ profile of HERAPDF fits with free $\alpha_s$, with 
and without the 
jet data, from which we can appreciate that $\alpha_S$ can only be determined 
with the additon of jet data. This figure also shows the PDFs from HERAPDF1.6
free $\alpha_s(M_Z)$ and the $q-\bar{q}$ and $g-g$ luminosity plots for the 
HERAPDF NLO updates of 2011.

\section{NNLO PDFs and predictions for the Higgs}
The Higgs cross section is very sensitive to the value of $\alpha_s(M_Z)$ and 
the gluon-gluon luminosity. For Higgs predictions it is also necessary to 
consider NNLO calculations. Until this year there were only a few NNLO 
PDFS~\cite{Alekhin:2010dd}. The PDF4LHC recommendation concentrated on the 
MSTW2008 PDF beacuse it is a global fit. However the predictions of the JR, 
ABKM and HERAPDF1.0 NNLO PDFs all lie to the low side of MSTW2008 and this 
was used~\cite{Baglio:2011wn} to cast doubt on the Higgs exclusion limits from the 
Tevatron which were based on MSTW. Two further issues have arisen related to 
this. Firstly, Accardi et al~\cite{Accardi:2011fa} have reconsidered deuterium 
corrections for the fixed target data. They find much larger 
corrections than have 
usually been accounted for, and this results in greater uncertainty in 
the high-$x$ $d-$quark which feeds into the high-$x$ gluon PDF 
when it is determined from 
Tevtaron jet data where the $d-g$ process provides a substantial part of the 
cross section. Secondly, ABM~\cite{Alekhin:2011ey} have examined the use 
of NMC data in the global 
fits. The NMC data on $F_2$ have often been used for the PDF fits rather than 
the cross section data. However, the extraction of $F_2$ relies on assumptions 
on the value of $F_L$ which may not be consistent with modern QCD caluclations.
ABM find that using $F_2$ rather than the cross section raises their extracted 
values of $\alpha_S$ erroneously. However both MSTW\cite{Thorne:2011kq} and NNPDF\cite{:2011we} have repeated 
this analysis and do not agree with these conclusions (at the time of writing 
this is still not resolved).

Since the gluon PDF is so important for the Higgs predictions another issue 
which has been raised is the goodness of fit of the ABKM and HERAPDFs to the 
Tevatron jet data. Watt and Thorne~\cite{Thorne:2011kq} obtain poor $\chi^2$ for these data when fitting to the  
HERAPDF1.0, 1.5 and the ABKM09 PDF predictions. However these fits only 
compare to the central predictions of the HERAPDF and ABKM PDFs. A more valid 
comparison would account for the PDF error bands. HERAPDF have input the 
Tevatron jet data to their fit and they obtain much better $\chi^2$
($\chi^2/ndp = 1.48$ for CDF and $1.35$ for $D0$ jets). 
Significantly the resulting PDFs do not lie outside the HERADF1.5 error bands
(although they do imply a harder high-$x$ gluon on the upper edge of the error
 band). ABKM have also made their own fits obtaining $\chi^2/ndp= 0.94$ for 
D0 di-jet data.

\begin{figure}[tbp]
\vspace{-2.0cm} 
\centerline{\psfig{figure=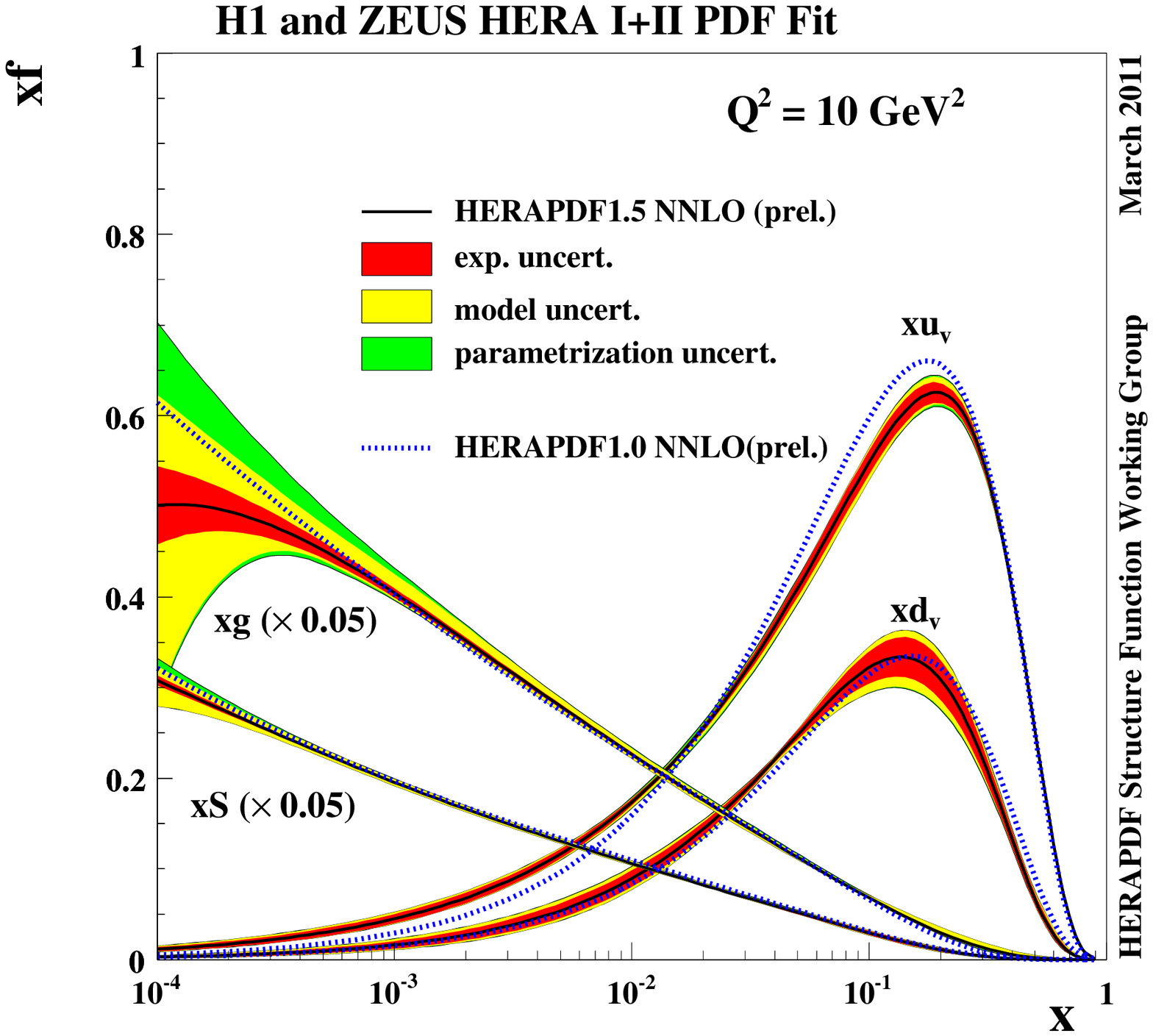,width=0.30\textwidth}~~ 
\psfig{figure=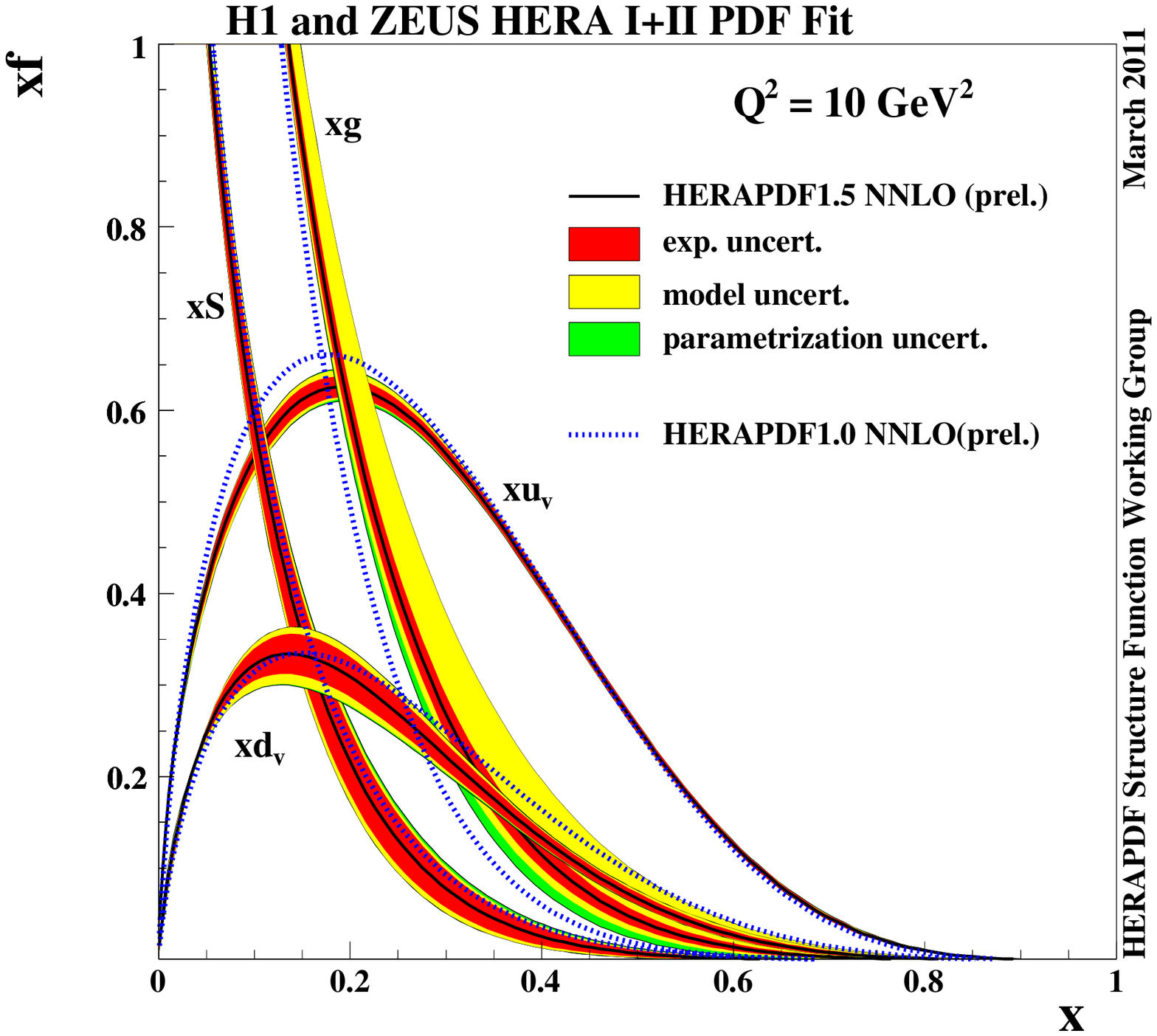,width=0.30\textwidth}}
\centerline{\psfig{figure=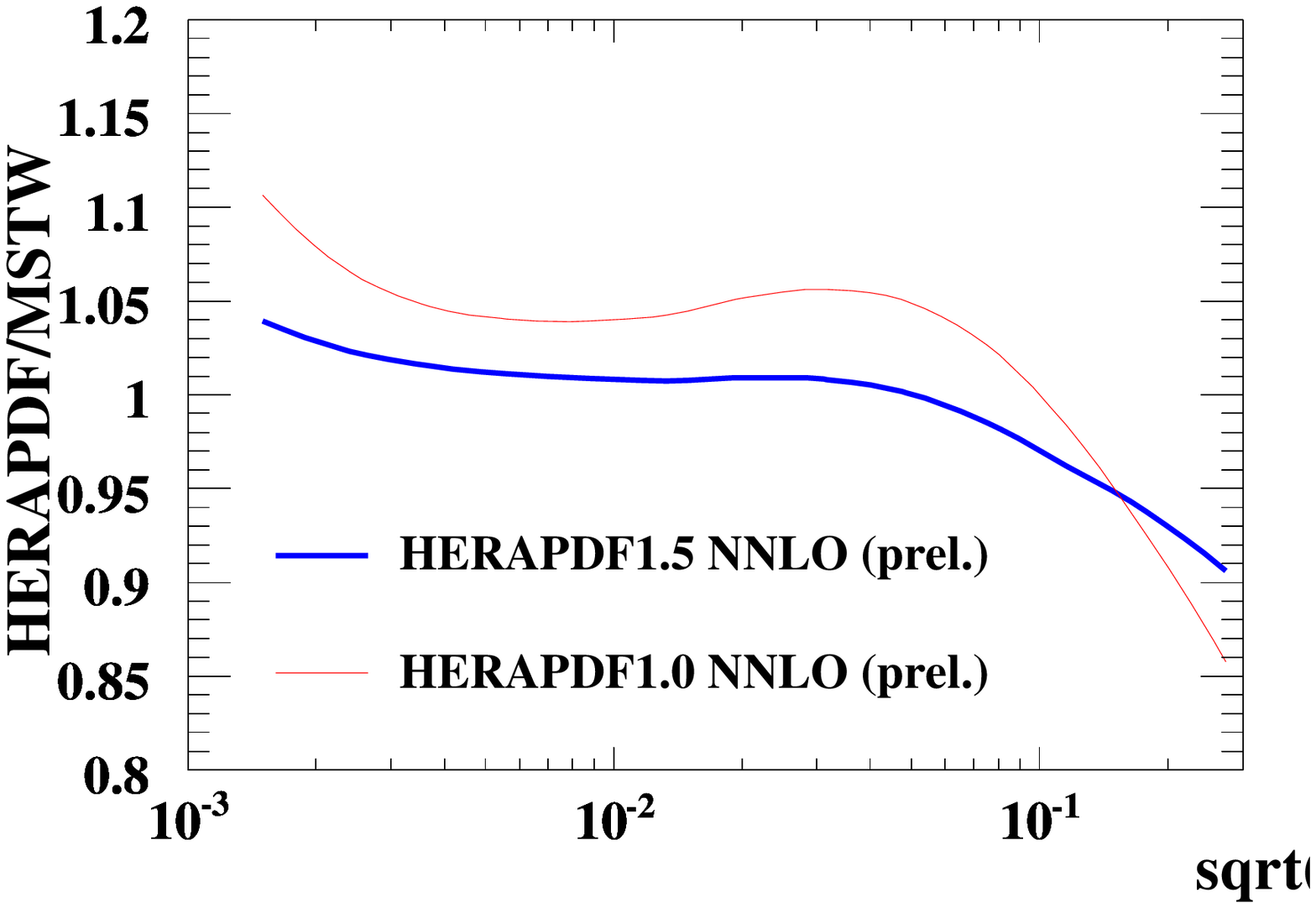,width=0.30\textwidth}~~ 
\psfig{figure=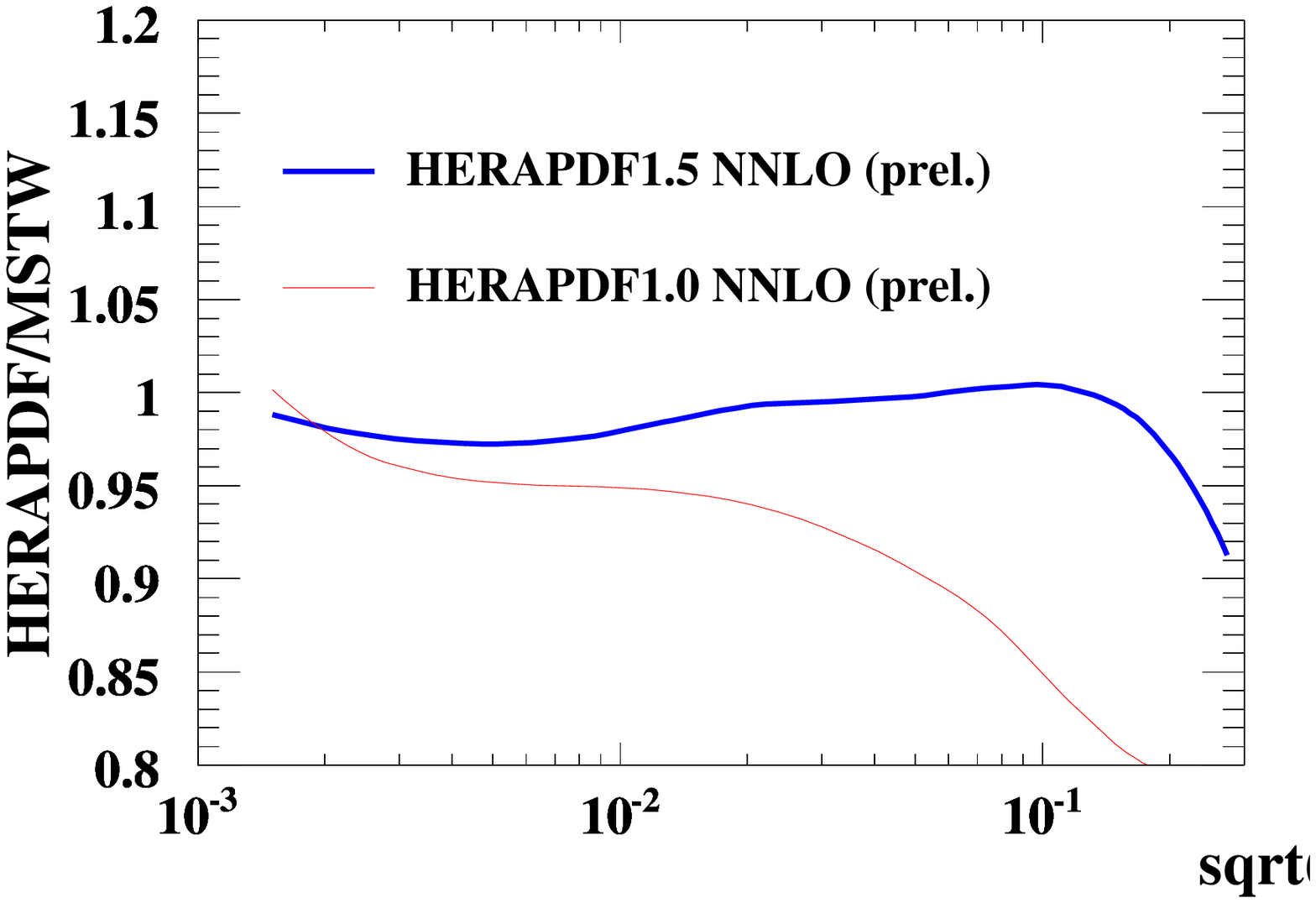,width=0.30\textwidth}}
\caption {Top row:HERAPDF1.5NNLO PDFs compared to HERAPDF1.0NNLO PDFS on log 
and linear $x$ scales. Bottom row: $q-\bar{q}$ and $g-g$ luminosity plots 
for HERAPDF1.5 and 1.0 NNLO in ratio to MSTW2008NNLO.
}
\label{fig:nnloherapdf}
\end{figure}

In 2011 many more NNLO PDFs are becoming available. 
NNPDF presented a preliminary NNLO analysis called NNPDF2.5 
(which will be called NNPDF2.1 NNLO) and HERAPDF presented a new 
NNLO extraction 
HERAPDF1.5NNLO~\cite{herapdf15nnlo}, with full accounting for experimental, model and 
parametrisation uncertainties, using the extended form of their 
parametrisation. An NNLO PDF set from the CT group is also expected soon.
Scale differences in NNLO heavy quark calculations are significantly reduced 
such that NNLO predictions should be in better agreement regardless of the 
choices made for these schemes. 
Fig.~\ref{fig:nnloherapdf} compares the 
HERAPDF1.5NNLO with the preliminary HERAPDF1.0 NNLO which was issued only as 
central predictions for two values of $\alpha_s$. The HERAPDF1.5 NNLO fit 
has a harder high-x gluon. The figure also shows the $q-\bar{q}$ and $g-g$
HERAPDF NNLO luminosities in ratio to MSTW2008, illustrating much
closer agreement with MSTW2008 for HERAPDF1.5 than for HERAPDF1.0. This, added
to the fact that HERAPDF now recommend that a central 
value of $\alpha_s(M_Z)=0.1176$ be used at NNLO, brings the Higgs predictions 
from HERAPDF into much closer agreement with those of MSTW2008.

\section{Comparison to early LHC data}
\begin{figure}[tbp]
\vspace{-3.0cm} 
\centerline{\psfig{figure=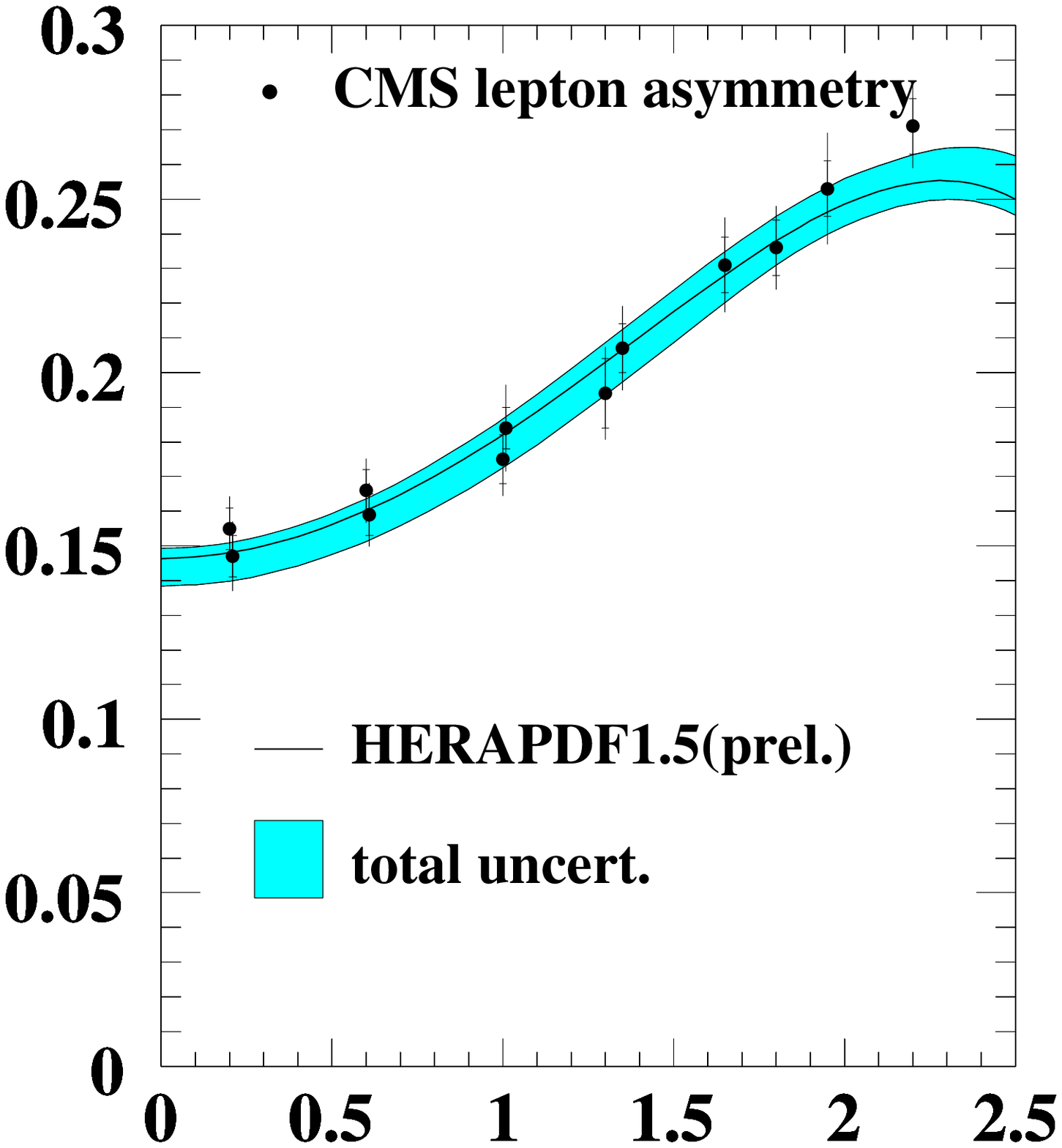,width=0.30\textwidth}~~ 
\psfig{figure=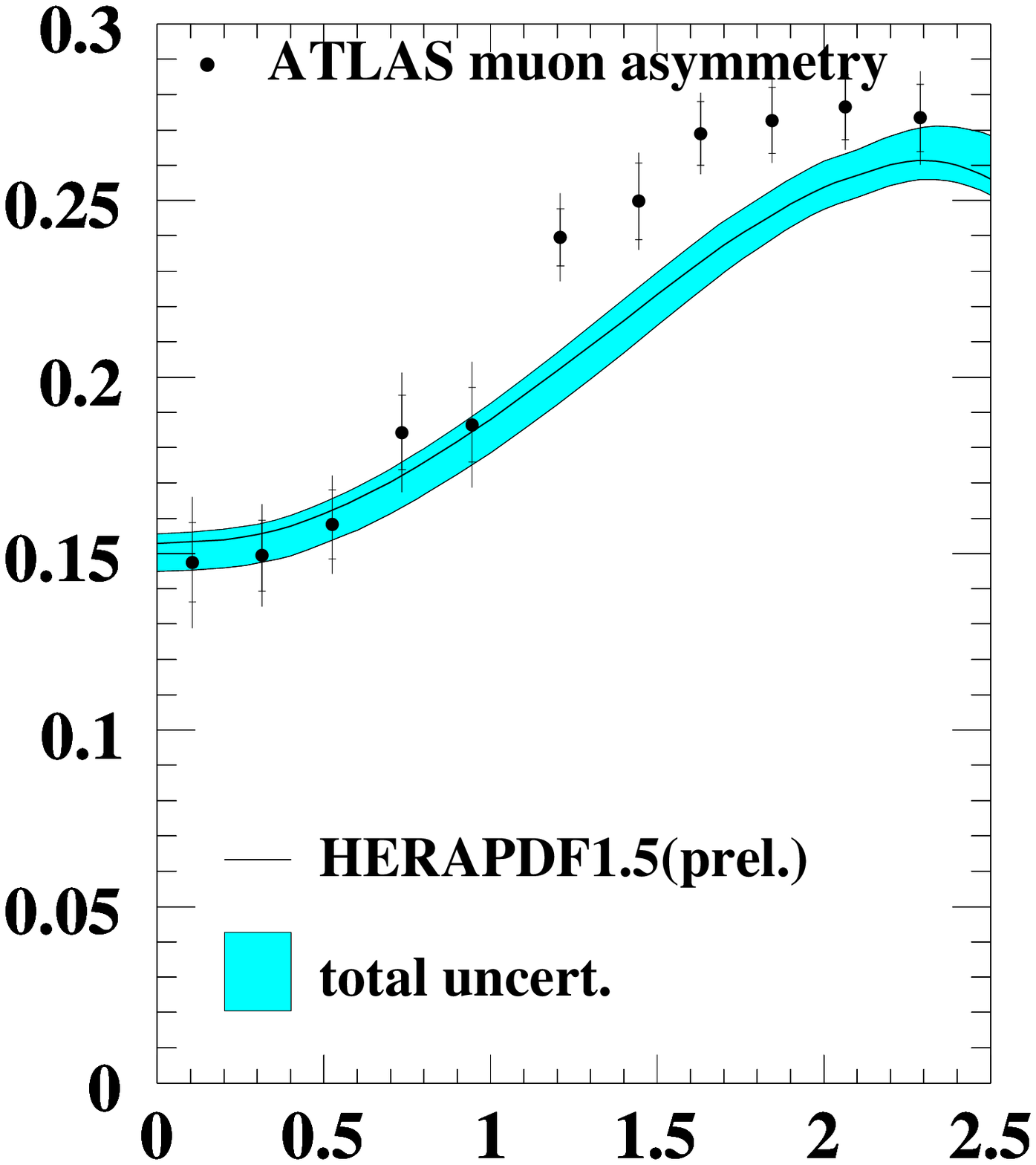,width=0.30\textwidth}}
\centerline{\psfig{figure=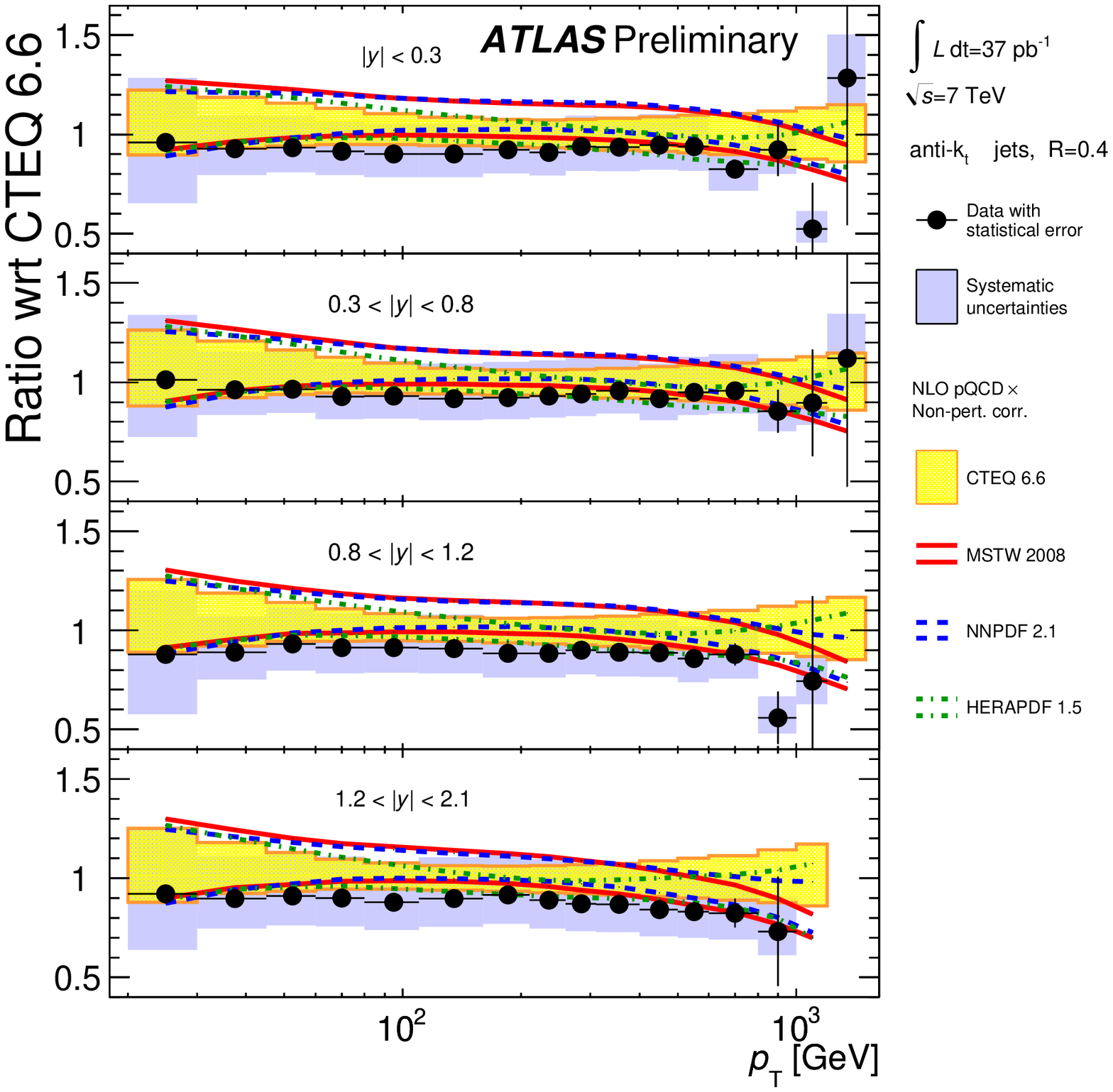,width=0.30\textwidth}~~ 
\psfig{figure=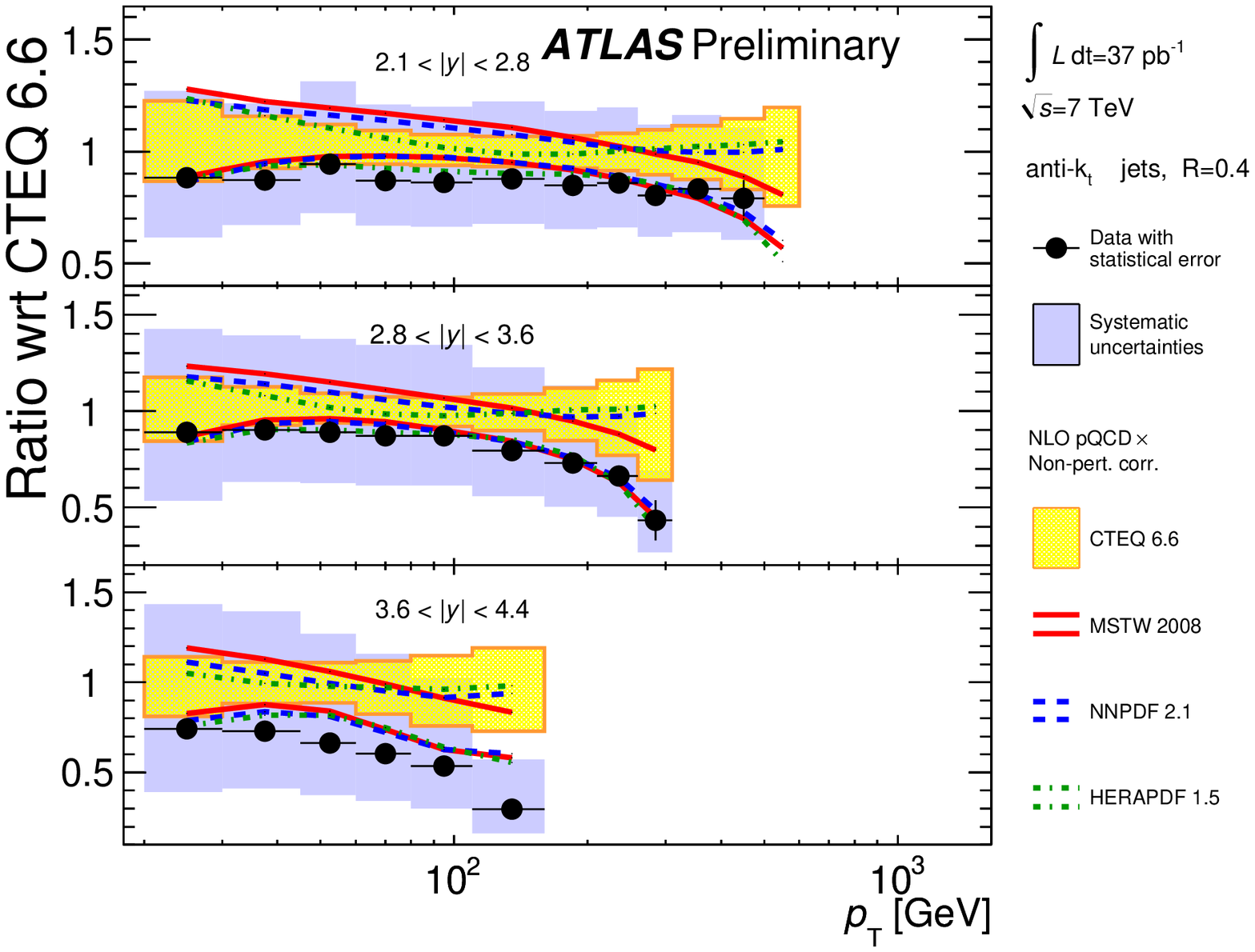,width=0.30\textwidth}}
\caption {Top: HERAPDF1.5 predictions for LHC $W$-lepton asymmetry data from CMS and ATLAS data. Bottom ATLAS jet data in the central and forward regions in ratio to the predictions of CTEQ6.6 and compared to other PDF predictions.
}
\label{fig:LHC}
\end{figure}

Fig.~\ref{fig:LHC} show comparisons of the HERAPDF1.5 NLO
 predictions to the early 
LHC data on the W asymmetry from CMS and ATLAS. Predictions from other PDFs show a similar level of agreement. This figure also show comparisons of various 
PDFs to the ATLAS inclusive jet data. 
The W data have also 
been used for very preliminary evaluations of their impact on the PDF 
uncertainties by both the HERAPDF and the NNPDF groups. 
NNPDF~\cite{Ball:2010gb} have used a reweighting technique
 to asses the 
impact of these data, finding improvements of the order of $\sim 40\%$ in the 
u and d-quark. HERAPDF have fitted the asymmetry data from CMS in addition to 
the inclusive HERA data and the CDF Z0 and W-asymmetry data. They find a 
modest improvement for the valence quark densities at low-$x$ - the region 
that Tevatron data do not reach.

Plots illustrating these improvements and further comparisons of the PDFs 
to each other and to Tevatron and early LHC data can be found at

https://wiki.bnl.gov/conferences/images/1/1a/Plenary.Cooper-Sarkar.0415.talk.pdf 

See also the review of Watt~\cite{Watt:2011kp} 
for recent work on benchmarking the PDFs.


\begin{footnotesize}

\end{footnotesize}


\end{document}